\RequirePackage{fix-cm}
\documentclass[smallextended]{svjour3}       
\smartqed  
\usepackage{amsmath}
\allowdisplaybreaks

\usepackage{graphicx}
\usepackage{subfigure}

\newcommand{\Frac}[2]{\mbox{$\displaystyle\frac{#1}{#2}$}}

\usepackage{xcolor}
\definecolor{newcolor}{rgb}{.8,.349,.1}
\begin{document}

\title{Analysis of the orbital evolution of space debris using a solar sail and natural forces 
}


\author{Jean Paulo dos Santos Carvalho         \and
        Rodolpho Vilhena de Moraes \and
        Antonio Fernando Bertachini de Almeida Prado
}


\institute{J. P. S. Carvalho \at
              Science and Technology Center in Energy and Sustainability, Federal University of the Rec\^{o}ncavo of the Bahia, Av. Centen\'{a}rio 697, Feira de Santana-BA 44042-280, Brazil \\
              \email{jeanfeg@gmail.com}           
           \and
           R. V. de Moraes \at
              Institute of Science and Technology, Federal University of S\~{a}o Paulo, Av. Cesare Mansueto Giulio Lattes 1201, S\~{a}o Jos\'{e} dos Campos - SP, 12.247-014, Brazil
              \and
           A. F. B. A. Prado \at
           Division of Space Mechanics and Control, National Institute for Space Research, Av. dos Astronautas, 1.758, S\~{a}o Jos\'{e} dos Campos, SP, Brazil
}

\date{Received: date / Accepted: date}

\maketitle

\begin{abstract}
Since the launch of the first artificial satellite around the Earth up to the present date, various non-functional objects are orbiting the Earth, such as deactivated satellites, satellite components, rocket bodies, among others. These objects can collide with operational satellites or other debris, generating a cloud of smaller particles. Space agencies are concerned about the current scenario of the space environment. If we continue as we are the number of objects in orbit will make it difficult to operate safely in the space environment. So space exploration may become unsustainable (\cite{ESA}). In this work, the orbital evolution of these objects that are located in the geostationary orbit (GEO) is analyzed. In the mathematical model we consider the main disturbing forces, such as the direct solar radiation pressure (SRP), the perturbation of the third body (Sun and Moon) and the Earth oblateness. The SRP is the most relevant perturbation for objects with larger area-mass ratio. Knowing this, the possibility of using a solar sail is considered to help to clean the space environment. The main natural environmental perturbations that act in the orbit of the debris are considered in the dynamics. Such forces acting in the solar sail can force the growth of the eccentricity of these objects in the GEO orbit. Several authors have presented models of the solar radiation pressure considering the single-averaged model. But, doing a literature research, we found that the authors consider the Earth around the Sun in a circular and inclined orbit. Our contribution to the SRP model is in developing a different approach from other authors, where we consider the Sun in an elliptical and inclined orbit, which is valid for other bodies in the solar system when the eccentricity cannot be neglected. The expression of the SRP is developed up to the second order. We found that the first-order term is much superior to the second-order term, so the quadrupole term can be neglected. Another contribution is the approach to identify the initial conditions of the perigee argument ($g$) and the longitude of the ascending node ($h$), where some values of the ($g, h$) plane contribute to amplify the eccentricity growth. In the numerical simulations we consider real data from space debris removed from the site Stuff in Space. The solar sail helps to clean up the space environment using a propulsion system that uses the Sun itself, a clean and abundant energy source, unlike chemical propellants, to contribute to the sustainability of space exploration.

\keywords{Space debris \and Solar radiation pressure \and Solar sail \and Sustainability \and Astrodynamics}
\end{abstract}

\section{Introduction}
\label{intro}
Orbiting the Earth there are artificial satellites operating for various services that are used by our society, such as, telecommunications, navigation, geolocation, weather forecasts and space research. However, since the beginning of space exploration, the space environment has been mostly populated by non-functional objects of different sizes. These objects are known as space debris or simply space junk. Space debris are all artificial objects, including fragments and their elements, in Earth orbit or re-entering the atmosphere, which are not functional. The debris cause space pollution and, in the event of a collision, pose risks to space exploration and active satellites in orbit. According to \cite{ESA}, only about $6\%$ of the total number of known man-made objects in space, larger than 10 centimeters, are operational. Because of this obstacle, there is an interest in seeking solutions to mitigate the problem of the space debris. In this paper, the use of solar sail as a mitigation technique is analyzed, which are objects with a large surface area and small mass, which allows movement using the solar radiation pressure as a source of propulsion to mitigate the problem of the space debris. The modeling and exploration of the SRP is widely discussed and applied in the literature, such as \cite{Ferraz,Lucking,Luckingb,Gkolias,Colombob,Colombo,Krivov,Krivovb,Xin}. In addition, a large amount of previous work has focused on the use of solar sails for end-of-life disposal, such as \cite{Krivov,Krivovb,Colombo}. In this way, we present an approach that can contribute to the sustainability of the space exploration, using the solar sail as a propulsion mechanism. We analyzed the orbital evolution of the debris in the geostationary orbit (GEO). The main disturbing forces considered in the dynamics are the direct solar radiation pressure (SRP), the perturbation of the third body (PTB) and the oblateness of the Earth. The SRP is the most relevant disturbance for objects with a large area-to-mass ratio (\cite{Moraes,Casanova}). We emphasize that we do not consider the effect of the Earth shadow on debris, as done in \cite{Hubaux} (see \cite{Ferraz} for a shadow function model). This effect does not significantly contribute to the dynamics, as commented in \cite{Gachet}. Much of the works on space debris neglects the shadow of the Earth, but in \cite{Fruh}, this effect is considered in the dynamics of objects with high area-to-mass ratios. Where Self-shading methods have been developed.

The main goal of this work is to study the use of a solar sail to contribute with the increase of the eccentricity of the debris, in particular, using the SRP. It is known that in GEO orbit the eccentricity is almost zero with no larger variations. So, according to other works, \cite{Lucking,Luckingb,Colombob,Valkb,Colomboc} it is shown that a very large area-to-mass to deorbit a satellite is required in GEO. Such values are beyond current and near-future technologies of solar sails. As the technology to obtain A/m ratios of the order of 20-30 $m^2/kg$ still has its limitations, in this work, we present a study of the dynamics of a supposed solar sail, highlighting that these values (including 40 m$^2$/kg) appear in several works in the literature (\cite{Casanova,Maria,Colombo,Valk}). Thus, by attaching a solar sail to the debris, it can approach the surface of the Earth in a much shorter time than using natural decay - which can take hundreds of years. As shown in \cite{Oltroggea}, the knowledge of the collisions probability for satellites operating in geosynchronous Earth orbit (GEO) is of extreme importance and interest to the global community and GEO spacecraft operators. See, for example, \cite{Colombod,Colomboe}, where the effect of solar sails to the collision probability during re-entry was analysed, or in \cite{Gkolias}, where it was verified that the risk of collision for spacecraft deorbiting from highly inclined GEO without a sail is small. Some papers motivated the development of this work (\cite{Casanova,Gkolias,Elisa,Rossi}). For example, according to \cite{Elisa}, one of the fundamental issues to mitigate the problem of the space debris is to check for the existence of natural perturbations that facilitate the reentry of this debris on Earth, to be destroyed by the effect of the atmospheric drag. In \cite{Casanova}, an analytical and numerical model is presented to propagate space debris in the GEO orbit. In \cite{Gkolias} the orbital dynamics of artificial satellites around the Earth at geosynchronous altitude is explored with the main objective of evaluating the current mitigation guideline, as well as discussing the future exploration of the region. At the end of its useful life the satellite must be maneuvered far enough away from GEO, into a graveyard orbit, to avoid interfering with spacecraft still operating in geostationary orbit. This maneuver must place the object in an orbit that remains above the region protected by the GEO. However, from a sustainability standpoint, the use of graveyard orbits, as suggested by current mitigation guidelines, will continue to increase the probability of collision in GEO (see \cite{Gkolias}). In \cite{Elisa}, the effect of the orbital resonances associated with the solar radiation pressure for low Earth orbit is analyzed. The authors have shown that at least four of the six resonances analyzed can be considered to increase the eccentricity of a small spacecraft equipped with a solar sail. In \cite{Alessandra} the secular resonance effect was used to show that the growth in the eccentricity, as observed in space debris located in the MEO region at inclination approximately equal to 56 degrees, can be explained as a natural effect of the secular resonance $2\dot{w}+\dot{\Omega}=0$. In \cite{Maria}, an analytical model to characterize the equilibrium points and the phase space associated with the single averaged dynamics caused by the non-sphericity of the Earth coupled with the perturbations of the solar radiation pressure is developed. The authors show the resonant curves that always act as separators between libration and circulation motion, when there is an equilibrium point with low eccentricity. The authors show that, for the same combination ($a$, $e$, $i$, $A/m$), there are equilibrium points associated with different resonant terms. In \cite{Sampaio}, the authors consider different perturbations and resonances in the orbital motions of space debris distributed at different altitudes. Objects in resonant orbital motions are studied in low Earth orbits. Using the two line elements of NORAD, resonant angles and resonant periods associated with real motions are described, providing a more precise formation for the development of an analytical model that describes a certain resonance. In \cite{San} the authors present a method for the end-of-life disposal of spacecraft in Medium Earth Orbit (MEO). To analyze the study of the long-term dynamics of MEO objects from a space debris perspective, as the region is becoming increasingly populous, for example, due to the launch of the Galileo and Beidou constellations. In \cite{Gachet}, the authors review the long-term dynamics of the geostationary Earth orbits through the application of the canonical perturbation theory, where a Hamiltonian model was considered for all the main perturbations: geopotential in order and degree two, lunisolar perturbations and solar radiation pressure. Applications are made for space debris in GEO orbits. The dynamics of objects with a high area-to-mass ratio has also been explored by \cite{Rosengren}, where a new averaged formulation of the orbital evolution of these objects has been analyzed taking into account the solar radiation pressure, oblateness of the Earth, and lunisolar perturbation. The first-order averaged model is given in terms of Milankovitch's orbital elements. In \cite{Valk}, the expansion of the disturbing potential from averaged up to degree 1 in the Legendre polynomials is presented explicitly. Note that the authors assumed that the relative motion of the Sun around the Earth is circular, due to the small eccentricity of the Earth orbit. In \cite{schaus}, the results of a numerical evaluation of the natural reduction in lifetime in low-altitude Earth orbit are shown, due to the dynamical perturbations. The results were obtained with two orbit propagators, one of a semi-analytical nature and the second one using equations of motion without averages. The simulations for the two propagators were compared. Both use the solar radiation pressure and the secular terms of the geopotential, including the $J_{2}$, $J_{4}$ and $J_{6}$ terms. In \cite{Moraes} a semi-analytical theory is developed to study the joint effects of the direct solar radiation pressure and atmospheric drag on the orbit of an artificial satellite around the Earth. The solutions eliminate the spurious terms in the Brouwer and Hori theory.

In \cite{Tresaco}, the double averaged model is applied to expand the solar radiation pressure term up to the second order considering the Sun in elliptical and inclined orbit. In the double averaged model, the first order term is constant and does not affect the orbital elements of the debris. Here, in this work, we also consider the mathematical model of the solar radiation pressure taking into account the Sun in an elliptical and inclined orbit, but now we use the single averaging model. We verified that the first-order term is predominant in relation to the second-order term, where the latter can be neglected. In \cite{Casanova,Gkolias,Colombo,Gkoliasb,Krivov,Krivovb} the authors also consider the single averaged model, but in the PRS mathematical model they consider the Sun in a circular and inclined orbit. In this way, our model is slightly different from the models presented. We also show a comparison of the present research with the work of \cite{Gkolias}. We compared the two equations by making the eccentricity of the orbit of the Sun to be zero. As mentioned before, the main objective of this work is to use the solar sail to amplify the growth of the eccentricity and inclination of the space debris in a geostationary orbit.

\section{Mathematical model}

We consider the motion of an object (spacecraft or debris) around the Earth, taking into account the gravitational attraction of the third body (Sun and Moon), the solar radiation pressure, without the effect of the shadow, and the non-uniform distribution of mass of the Earth ($J_{2}$).

\subsection{Solar radiation pressure}

Let us consider a $Oxyz$ reference frame centered on the main body, Earth. The $Oxy$ plane coincides with the equator of the planet, the $x$-axis is defined by the intersection line of the equatorial plane of the main body and the orbital plane of the third body (see Figure 1 in \cite{Paulo}). It is assumed that the third body follows an elliptical and inclined orbit around the main body with semimajor axis $a_{\odot}$, eccentricity $e_{\odot}$, inclination $i_{\odot}$, argument of the perigee $g_{\odot}$, longitude of the ascending node $h_{\odot}$ and true anomaly $f_{\odot}$. The spacecraft orbits around the central body with semimajor axis $a$, eccentricity $e$, inclination $i$, argument of the perigee $g$, longitude of the ascending node $h$ and true anomaly $f$. Its motion is perturbed by the third body, $J_{2}$ and SRP. The equation of motion of the spacecraft is given by

\begin{equation}\label{1}
\begin{array}{l}
\ddot{\mathbf{r}} = \ddot{\mathbf{r}}_M +\ddot{\mathbf{r}}_{3b}+\ddot{\mathbf{r}}_{SRP},
\end{array}
\end{equation}
where $ \mathbf{\ddot{r}_M}$ is the force induced by Earth gravity field, which can be expressed as the gradient of a certain potential $U_{M}$. It is written in terms of the position vector $\mathbf{r}$ of the spacecraft with respect to the planet Earth,

\begin{equation}\label{2}
\begin{array}{l}
 \ddot{\mathbf{r}}_M= \nabla U_M(\mathbf{r}).
\end{array}
\end{equation}

The term  $\ddot{\mathbf{r}}_{3b}$  is the resultant of the gravitational attraction of the third body and can be referred to center of mass of the Earth

\begin{equation}\label{3}
\begin{array}{l}
 \ddot{\mathbf{r}}_{3b}= -\mu_{\odot}\left(\Frac{\mathbf{r}-\mathbf{r}_{\odot}}{\|\mathbf{r}-\mathbf{r}_{\odot}\|^3}+
 \Frac{\mathbf{r}_{\odot}}{\|\mathbf{r}_{\odot}\|^3}\right),
\end{array}
\end{equation}
where $\mu_{\odot}$ represents the gravitational parameter of the Sun and $\mathbf{r_{\odot}}$ is the position vector of the Sun with respect to Earth \textcolor{black}{(the same applies when the third body is the Moon)}.

The acceleration generated by the solar radiation pressure is named $\ddot{\mathbf{r}}_{SRP}$. Therefore, the solar sail acceleration is expressed as (\cite{Paulo})

\begin{equation}\label{4}
\begin{array}{l}
 \ddot{\mathbf{r}}_{SRP}= \Frac{L_\odot}{c\, 2\pi \rho^2}\,\Frac{A}{m}\,(\mathbf{u}_i\cdot \mathbf{n})^2\mathbf{n},
\end{array}
\end{equation}
where $m$ is the mass of the spacecraft, $L_\odot$ is the Sun's luminosity, $\rho$ is the average distance from the planet to the Sun, in the case of the Earth $\rho=1$ AU and $c$ the speed of light. We will assume that the sail keeps a fixed orientation perpendicular to the Sun line ($\mathbf{u}_i \| \mathbf{n}$), then the sail effect takes its maximum value when $(\mathbf{u}_i\cdot \mathbf{n})=1$. The solar sail acceleration, given in Eq. (\ref{4}), in terms of the area-to-mass ratio $A/m$ of the sail, is also expressed in terms of the lightness number of the sail: $\beta$ (see \cite{McInnes}). It is a dimensionless ratio of the solar radiation pressure acceleration to the solar gravitational acceleration that measures the effectiveness of the sail. $\beta$ is defined as

\begin{equation}\label{5}
\begin{array}{l}
 \beta=\Frac{\sigma^*}{\sigma},
\end{array}
\end{equation}
where

\begin{equation}\label{6}
\begin{array}{l}
\sigma^*=\Frac{L_\odot}{c\, 2\pi \mu_{\odot}},
\end{array}
\end{equation}
for the case of the Earth we get $\sigma^*=1.53$ $gr/m^2$ (\cite{McInnes}). Here $\sigma$ is the sail loading parameter (areal density) given by the total mass of the spacecraft divided by the sail area, $\sigma={m}/{A}$, expressed in $gr/m^2$. Thus, taking into account that the sail is assumed to maintain the orientation perpendicular to the Sun line $\mathbf{n}=\mathbf{ \rho}/\|\mathbf{\rho}\|$, and the sail is located at a distance from the Sun $\mathbf{\rho}=\|\mathbf{r}-\mathbf{r}_{\odot}\|$, the solar radiation pressure is given by (\cite{Paulo})

\begin{equation}\label{7}
\begin{array}{l}
\mathbf{\ddot{r}}_{SRP}= \beta\,\mu_{\odot}\Frac{\mathbf{r}-\mathbf{r_{\odot}}}{\|\mathbf{r}-\mathbf{r_{\odot}}\|^3}.
\end{array}
\end{equation}

As commented in some works (\cite{McInnes,Paulo}), the force on a solar sail depends on the sail area and orientation. When the perturbation term due to solar radiation pressure  is added to the system, it only holds its Hamiltonian character when the sail is aligned with respect to the Sun, i.e., no sail effect, or in the case that the sail is perpendicular to the Sun-sail direction, i.e., maximum sail effect. \textcolor{black}{In \cite{Miguel}, the authors study desorbitation using an analogue to the quasi-rhombic-pyramid concept for planar motion. The focus of the work was to maintain a stable attitude close to the direction of the velocity of the spacecraft relative to the atmosphere. The effect of the $J_{2}$ term, SRP and atmospheric drag was considered. The expressions for the accelerations and torques due to the SRP and drag were derived in \cite{Miguelc}. For other sail orientations the systems are no longer Hamiltonian. The work of \cite{Miguel} (see also \cite{Miguelb}) analysed the problem of a sail with different attitude orientation and when the system could be treated still as Hamiltonian. It is known that the perturbation due to solar radiation pressure strongly depends on the size, mass and altitude of the orbit of the spacecraft, therefore varying the area-to-mass coefficient of the satellite and its altitude in order to visualize its impact on the sail dynamics.} According to \cite{Paulo}, considering the solar radiation acceleration, if we assume the case of a solar sail always perpendicular to the Sun-sail direction, the Sun gravitational attraction and the sail acceleration  are equal in magnitude (except for the scaling effect of $\beta$) but in opposite directions. From Eq. (\ref{7}) we get that the sail acceleration due to the radiation pressure can be expressed as the gradient of the following potential

\begin{equation}\label{8}
\begin{array}{l}
U_{SRP}=-\beta\mu_{\odot}\left(\Frac{1}{\|\mathbf{r}-\mathbf{r_{\odot}}\|}\right)
\end{array}
\end{equation}

Since we are interested in the case $r\ll r_{\odot}$, the term is expanded as a series of Legendre polynomials up to the second order,

\begin{equation}\label{10}
\begin{array}{l}
\Frac{1}{\|\mathbf{r}-\mathbf{r}_{\odot}\|}=\Frac{1}{r_{\odot}}\left(1-2\Frac{r}{r_{\odot}}\cos\psi+
\big(\Frac{r}{r_{\odot}}\big)^2\right)^{-1/2}=\\[1.5ex]
\qquad\Frac{1}{r_{\odot}}\sum_{n\geq0}^2\left(\Frac{r}{r_{\odot}}\right)^nP_n(\cos\psi).
\end{array}
\end{equation}

This potential function can be expanded in terms of Legendre polynomials up to the second order. We get:

\begin{equation}\label{9}
\begin{array}{l}
U_{SRP}=-\beta\Frac{\mu_{\odot}}{r_{\odot}}\left[1+\Frac{r}{r_{\odot}}\cos\psi+
  \Frac{1}{2} \left(\Frac{r}{r_{\odot}}\right)^2 (3\cos^2\psi-1)  \right],
\end{array}
\end{equation}
the angle $\psi$ is the angle between the radius vectors $\mathbf{r}$ and $\mathbf{r_{\odot}}$, which can be expressed in terms of the orbital elements of the Sun and the spacecraft.

Note that the first term of Eq. (\ref{9}) is constant, it does not depend on the spacecraft position. In \cite{Paulo} the double averaged model is applied, where the first-order term of Eq. (\ref{9}) is also constant, and thus, only the second-order term of the Legendre polynomial affects the orbit of the space vehicle. In the present research, we show an expansion of the SRP model presented in \cite{Paulo}. Now, the single averaged model is applied to develop the SRP equation. In this model, the second and third order terms of Eq. (\ref{9}) depend on the position of the spacecraft. We present here only the first order term, we get

\begin{equation}\label{10}
\begin{array}{l}
U_{SRP}=-\beta\Frac{\mu_{\odot}}{r_{\odot}}\left[\Frac{r}{r_{\odot}}\cos\psi\right],
\end{array}
\end{equation}
now, multiplying Eq. (\ref{10}) by $\left (a_{\odot}/a_{\odot}\right)^{2}$ and $ (a/a)$, we get



\begin{equation}\label{11}
\begin{array}{l}
U_{SRP}= \Frac{-\beta \mu_{\odot} a}{a_{\odot}}\Frac{{a_{\odot}}^{2}}{r_{\odot}^{2}}\Frac{{r}}{a}\cos( \psi).
\end{array}
\end{equation}

Using equation (4) of \cite{Tadashi} for the expression of $\cos(\psi)$, we get

\begin{equation}\label{12}
\begin{array}{l}
\cos(\psi)=C\cos ( f+g+h-{ f_{\odot}}-{ g_{\odot}}-{ h_{\odot}}) +\\[1.5ex]
\qquad A\cos( f+g+h+{ f_{\odot}}+{g_{\odot}}-{h_{\odot}}) +\\[1.5ex]
\qquad B\cos( f+g-h+{f_{\odot}}+{g_{\odot}}+{h_{\odot}}) +\\[1.5ex]
\qquad D\cos( f+g-h-{f_{\odot}}-{g_{\odot}}+{h_{\odot}
}) +\\[1.5ex]
\qquad E ( \cos( f+g-{ f_{\odot}}-{g_{\odot}}) -\cos( f+g+{ f_{\odot}}+{g_{\odot}})),
\end{array}
\end{equation}
where

$A=1/4\, \left( 1+\cos \left( i \right)  \right)  \left( 1-\cos \left(
{\it i_{\odot}} \right)  \right)$;

$B=1/4\, \left( 1-\cos \left( i \right)  \right)  \left( 1+\cos \left(
{\it i_{\odot}} \right)  \right)$;

$C=1/4\, \left( 1+\cos \left( i \right)  \right)  \left( 1+\cos \left(
{\it i_{\odot}} \right)  \right)$;

$D=1/4\, \left( 1-\cos \left( i \right)  \right)  \left( 1-\cos \left(
{\it i_{\odot}} \right)  \right)$;

$E=1/2\,\sin \left( i \right) \sin \left( {\it i_{\odot}} \right) $.

We made a change of variable to perform the average over the true anomaly ($f$) of the spacecraft, where a change in the integration variable is adopted for eccentric anomaly ($\nu$). This is done by using known equations from the celestial mechanics, which are:

\begin{equation}\label{13}
\begin{array}{l}
\sin(f)=(\sqrt{1-e^2}\sin(\nu))/(1-e\cos(\nu));
\end{array}
\end{equation}

\begin{equation}\label{14}
\begin{array}{l}
\cos(f)=(\cos(\nu)-e)/(1-e\cos(\nu))
\end{array}
\end{equation}

\begin{equation}\label{15}
\begin{array}{l}
r/a=1-e\cos(\nu)
\end{array}
\end{equation}

\begin{equation}\label{16}
\begin{array}{l}
dl=(1-e\cos(\nu))d\nu.
\end{array}
\end{equation}

Now replacing Eqs. (\ref{12}), (\ref{13}), (\ref{14}) and (\ref{15}) in Eq. (\ref{11}), the average is made using Eq. (\ref{16}). After algebraic manipulations we get,

\begin{equation}\label{17}
\begin{array}{l}
U_{SRP}=\frac{3}{2}\beta \frac {\mu_{\odot}( 1+e_{\odot} \cos ( {f_{\odot}} ))^{2} a e}{{{a_{\odot}}}^{2}
(1- {e_{\odot}}^{2}) ^{2}}( ( \cos( {f_{\odot}}) \cos (
g-h) ( D) +\\[1.5ex]
\qquad \sin ( {f_{\odot}}) \sin
( g-h )( D) +A ( \cos ( { f_{\odot}}) \cos ( g+h ) -\\[1.5ex]
\qquad \sin ( {f_{\odot}}) \sin
( g+h))) \cos( { g_{\odot}}-{ h_{\odot}}) +\\[1.5ex]
\qquad ( -\sin( {f_{\odot}}) \cos( g-h) ( D) +\\[1.5ex]
\qquad \cos ( {f_{\odot}}) \sin( g
-h ) ( D) -\\[1.5ex]
\qquad A ( \cos ( {f_{\odot}})
\sin ( g+h) + \\[1.5ex]
\qquad
\sin ( {f_{\odot}} ) \cos ( g+h))) \sin ( {g_{\odot}}-{h_{\odot}}) +\\[1.5ex]
\qquad B
( \cos ( {f_{\odot}}) \cos ( {g_{\odot}}+{h_{\odot}}) -\\[1.5ex]
\qquad \sin( {f_{\odot}}) \sin( {g_{\odot}}+{h_{\odot}})) \cos( g-h) -\\[1.5ex]
\qquad B( \cos( {
f_{\odot}}) \sin ( {g_{\odot}}+{h_{\odot}}) +\\[1.5ex]
\qquad \sin ( {f_{\odot}
}) \cos ( {g_{\odot}}+{h_{\odot}}) ) \sin ( g
-h) +\\[1.5ex]
\qquad C ( \cos ( {f_{\odot}} ) \cos( g+h) +\\[1.5ex]
\qquad \sin( {f_{\odot}}) \sin ( g+h)
) \cos ( {g_{\odot}}+{h_{\odot}}) +\\[1.5ex]
\qquad C( \cos( {f_{\odot}}) \sin ( g+h) -\\[1.5ex]
\qquad \sin ( {f_{\odot}})
\cos ( g+h)) \sin ( { g_{\odot}}+{h_{\odot}}
) +\\[1.5ex]
\qquad 2\sin ( g ) E ( \cos ( {f_{\odot}}) \sin ( {g_{\odot}}) +\\[1.5ex]
\qquad \cos ( {g_{\odot}} )
\sin ( {f_{\odot}} )) ),
\end{array}
\end{equation}
where ${{\frac {a_{\odot}}{{r_{\odot}}}}}$ has been replaced by

\begin{equation}\label{18}
\begin{array}{l}
{{\frac {a_{\odot}}{{r_{\odot}}}}}=(1+e_{\odot}\cos(f_{\odot}))/(1-e_{\odot}^2).
\end{array}
\end{equation}

Note that the single averaged equation is developed in a closed form for the orbital elements of the debris or spacecraft. Now, to replace the true anomaly of the disturbing body by the mean anomaly ($l_{\odot}$), we use the following expressions

\begin{equation}\label{19}
\begin{array}{l}
\sin{f_{\odot}}=\sin \left( {\it l_{\odot}} \right) +e_{\odot}\sin \left( 2\,{\it l_{\odot}} \right) +{e_{\odot}}^{2
} \left( {\frac {9\,\sin \left( 3\,{\it l_{\odot}} \right) }{8}}-{\frac {7\,
\sin \left( {\it l_{\odot}} \right) }{8}} \right)
\end{array}
\end{equation}

\begin{equation}\label{20}
\begin{array}{l}
\cos{f_{\odot}}=\cos \left( {\it l_{\odot}} \right) +e_{\odot} \left( \cos \left( 2\,{\it l_{\odot}}
 \right) -1 \right) +\\[1.5ex]
\qquad {e_{\odot}}^{2} \left( {\frac {9\,\cos \left( 3\,{\it l_{\odot}}
 \right) }{8}}-{\frac {9\,\cos \left( {\it l_{\odot}} \right) }{8}} \right),
\end{array}
\end{equation}
and after algebraic manipulations the disturbing potential of single averaged, with the Sun in elliptical and inclined orbit, is written in the form

\begin{figure}
\centering
\includegraphics[scale=0.4]{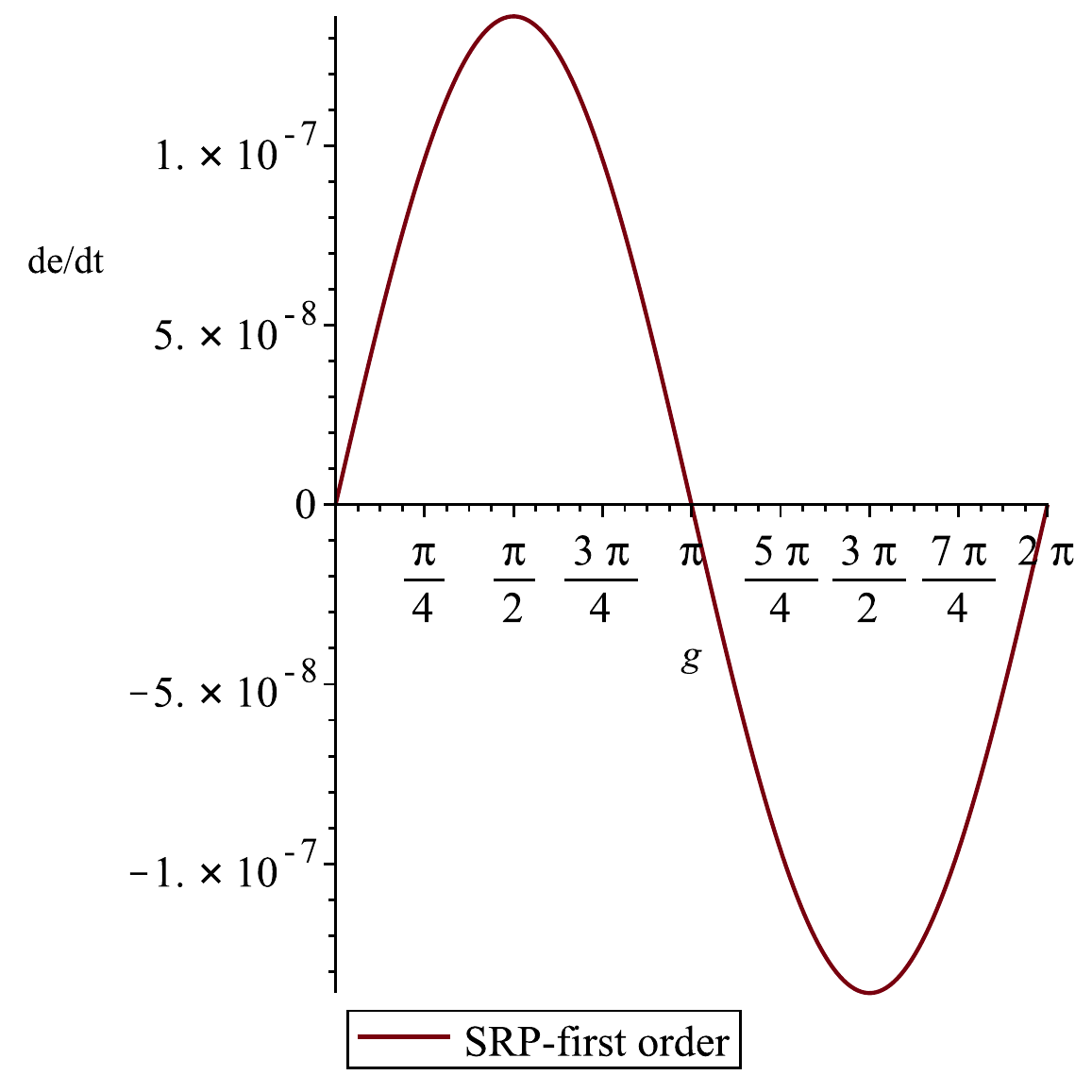}
\caption{E06321D debris. Initial conditions: $a=41400$ km, $e=0.035$, $i=7^{\circ}$, $g=0^{\circ}$, $h=0^{\circ}$ and $A/m=30$ $m^{2}/kg$. Disturbing potential: $R_{J2}+R_{SRP}+R2SA_{Sun}+R2SA_{Moon}$.}
\label{fig:T4}
\end{figure}

\begin{figure}
\centering
\includegraphics[scale=0.4]{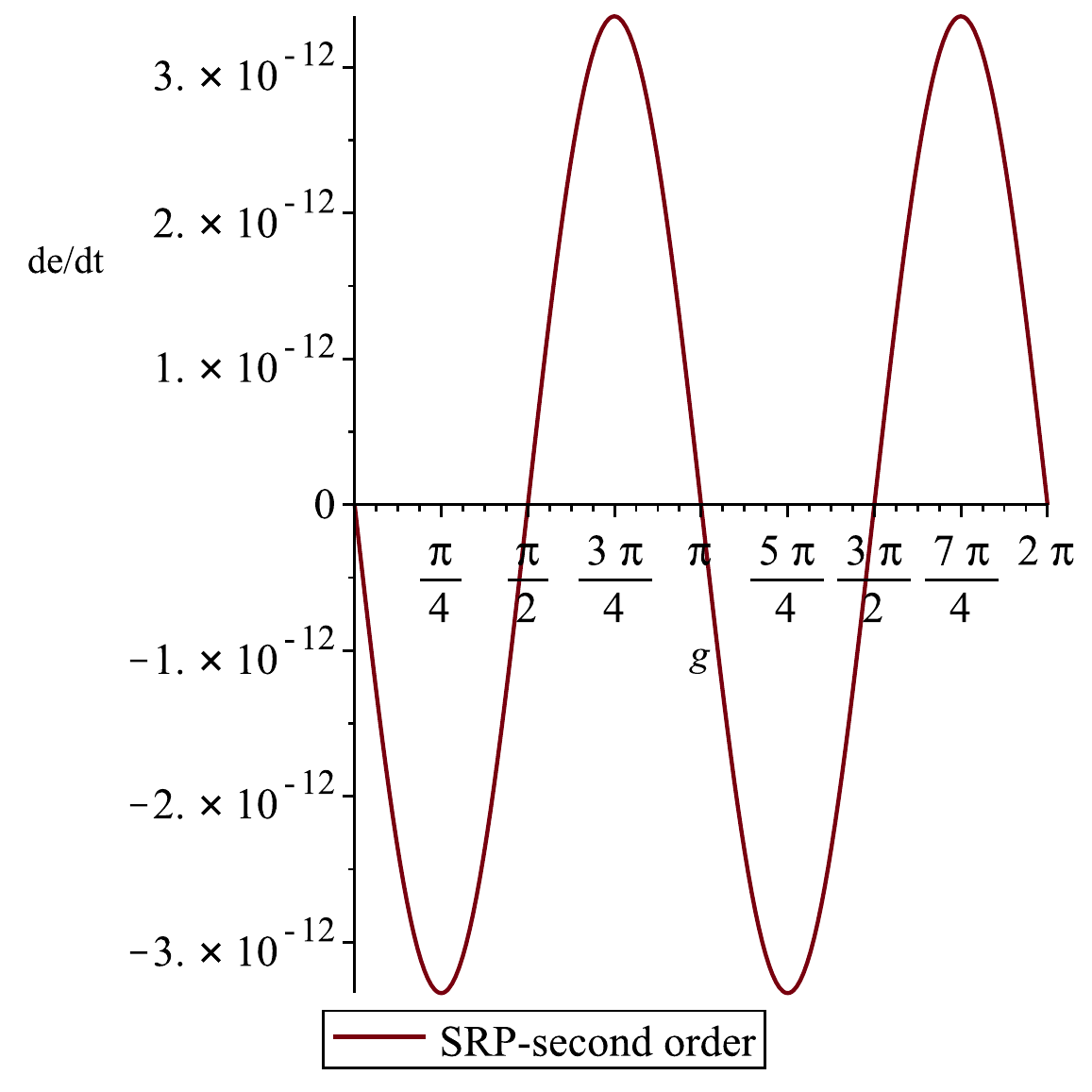}
\caption{E06321D debris. Initial conditions: $a=41400$ km, $e=0.035$, $i=7^{\circ}$, $g=0^{\circ}$, $h=0^{\circ}$ and $A/m=30$ $m^{2}/kg$. Disturbing potential: $R_{J2}+R_{SRP}+R2SA_{Sun}+R2SA_{Moon}$.}
\label{fig:T5}
\end{figure}

\begin{equation}\label{21}
\begin{array}{l}
<R_{SRP}>=\frac {81}{64} \beta \frac {e a {n_{\odot}}^{2} a_{\odot}}{( 1-{e_{\odot}}^{2})^{2}} \times \\[1.5ex]
\qquad ( -\frac{1}{27}  ( 20\cos  ( i  ) -20  )   (
\cos  ( {  i_{\odot}}  ) -1  )  \times \\[1.5ex]
\qquad
 ( {e_{\odot}}^{2}-2/5  ) \cos  ( -{  l_{\odot}}+g-h-{  g_{\odot}}+{  h_{\odot}}  ) -\\[1.5ex]
\qquad \frac{1}{27}  ( 20{
e_{\odot}}^{2}-8  )   ( \cos  ( i  ) +1  )   ( 1+
\cos  ( {  i_{\odot}}  )   )\times  \\[1.5ex]
\qquad \cos  ( -{  l_{\odot}}+g+h-{
g_{\odot}}-{  h_{\odot}}  ) +{e_{\odot}}^{2}\times \\[1.5ex]
\qquad  ( \cos  ( {  i_{\odot}}  ) -1
  )   ( \cos  ( i  ) -1  ) \cos  ( -3{
  l_{\odot}}+g-h-\\[1.5ex]
\qquad{  g_{\odot}}+{  h_{\odot}}  ) + {e_{\odot}}^{2}  ( \cos  ( i
  ) +1  )   ( 1+\cos  ( {  i_{\odot}}  )   )\times \\[1.5ex]
\qquad \cos  ( -3{  l_{\odot}}+g+h-{  g_{\odot}} {  h_{\odot}}  ) +\\[1.5ex]
\qquad \frac {16}{27} e_{\odot}
  ( \cos  ( {  i_{\odot}}  ) -1  )   ( \cos  ( i
  ) -1  )\times \\[1.5ex]
\qquad \cos  ( -2 {  l_{\odot}}+g-h-{  g_{\odot}}+{  h_{\odot}}
  ) +\\[1.5ex]
\qquad \frac {16}{27} e_{\odot}  ( \cos  ( i  ) +1  )
  ( 1+\cos  ( {  i_{\odot}}  )   )\times \\[1.5ex]
\qquad \cos  ( -2{
l_{\odot}}+g+h-{  g_{\odot}}-{  h_{\odot}}  ) -\\[1.5ex]
\qquad \frac{1}{27} {e_{\odot}}^{2}  ( 1+\cos
  ( {  i_{\odot}}  )   )   ( \cos  ( i  ) -1
  )\times \\[1.5ex]
\qquad \cos  ( -{  l_{\odot}}+g-h+{  g_{\odot}}+{  h_{\odot}}  ) -\\[1.5ex]
\qquad \frac{1}{27} {e_{\odot}
}^{2}  ( \cos  ( {  i_{\odot}}  ) -1  )  \times \\[1.5ex]
\qquad ( \cos
  ( i  ) +1  ) \cos  ( -{  l_{\odot}}+{  g_{\odot}}-{  h_{\odot}}+g
+h  ) +\\[1.5ex]
\qquad \frac{1}{27}{e_{\odot}}^{2}  ( \cos  ( {  i_{\odot}}  ) -1
  )   ( \cos  ( i  ) -1  ) \times \\[1.5ex]
\qquad \cos  ( {  l_{\odot}}
+g-h-{  g_{\odot}}+{  h_{\odot}}  ) +\\[1.5ex]
\qquad \frac{1}{27} {e_{\odot}}^{2}  ( \cos  ( i
  ) +1  )   ( 1+\cos  ( {  i_{\odot}}  )   ) \times \\[1.5ex]
\qquad
\cos  ( {  l_{\odot}}+g+h-{  g_{\odot}}-{  h_{\odot}}  ) -\\[1.5ex]
\qquad \frac {16}{27} e_{\odot}
  ( 1+\cos  ( {  i_{\odot}}  )   )   ( \cos  ( i
  ) -1  )\times \\[1.5ex]
\qquad \cos  ( 2{  l_{\odot}}+g-h+{  g_{\odot}}+{  h_{\odot}}
  ) -\\[1.5ex]
\qquad \frac {16}{27} e_{\odot}  ( \cos  ( {  i_{\odot}}  ) -1
  )   ( \cos  ( i  ) +1  ) \times \\[1.5ex]
\qquad \cos  ( 2{
l_{\odot}}+{  g_{\odot}}-{  h_{\odot}}+g+h  ) -\\[1.5ex]
\qquad {e_{\odot}}^{2}  ( 1+\cos  (
{  i_{\odot}}  )   )   ( \cos  ( i  ) -1  ) \times \\[1.5ex]
\qquad
\cos  ( 3{  l_{\odot}}+g-h+{  g_{\odot}}+{  h_{\odot}}  ) -\\[1.5ex]
\qquad {e_{\odot}}^{2}  (
\cos  ( {  i_{\odot}}  ) -1  )   ( \cos  ( i  )
+1  ) \times \\[1.5ex]
\qquad \cos  ( 3{  l_{\odot}}+{  g_{\odot}}-{  h_{\odot}}+g+h  ) +\\[1.5ex]
\qquad \frac{1}{27}
  ( 20\cos  ( i  ) -20  )   ( {e_{\odot}}^{2}-2/5
  )   ( 1+\cos  ( {  i_{\odot}}  )   )\times \\[1.5ex]
\qquad \cos  ( {
  l_{\odot}}+g-h+{  g_{\odot}}+{  h_{\odot}}  ) +  \frac{1}{27}  ( 20 \cos  ( {
  i_{\odot}}  ) -20  ) \times \\[1.5ex]
\qquad  ( {e_{\odot}}^{2}-2/5  )   ( \cos
  ( i  ) +1  )  \cos  ( {  l_{\odot}}+{  g_{\odot}}-{  h_{\odot}}+g+
h  ) + \\[1.5ex]
\qquad 2  (   ( -{\frac {20{e_{\odot}}^{2}}{27}}+{\frac{8}{27}}
  ) \cos  ( -{  l_{\odot}}-{  g_{\odot}}+g  ) +\\[1.5ex]
\qquad \cos  ( -3{
  l_{\odot}}-{  g_{\odot}}+g  ) {e_{\odot}}^{2}+ \frac {16}{27} \cos  ( -2{  l_{\odot}
}-\\[1.5ex]
\qquad {  g_{\odot}}+g  ) e_{\odot}-1/27{e_{\odot}}^{2}\cos  ( -{  l_{\odot}}+{
g_{\odot}}+g  ) +\\[1.5ex]
\qquad \frac{1}{27} {e_{\odot}}^{2}\cos  ( {  l_{\odot}}-{  g_{\odot}}+g  ) -
\frac {16}{27} e_{\odot}\cos  ( 2{  l_{\odot}}+{  g_{\odot}}+g  )-\\[1.5ex]
\qquad{e_{\odot}}^{2}
\cos  ( 3{  l_{\odot}}+{  g_{\odot}}+g  ) + \cos  ( {  l_{\odot}}+{
g_{\odot}}+g  )  \times \\[1.5ex]
\qquad  ( {\frac {20{e_{\odot}}^{2}}{27}}-{\frac{8}{27}}
  )   ) \sin ( i  ) \sin  ( {  i_{\odot}}  )
  ),
\end{array}
\end{equation}
here $n_{\odot}$ is the mean motion of the disturbing body. To test the order of magnitude due to the force of the solar radiation pressure, we replace the expressions of first (Eq. (\ref{21})) and second order in the Lagrange planetary equations and plot $de/dt$ with respect to $g$ (see Figs. \ref{fig:T4} and \ref{fig:T5}). We found that the effect of the first order term of the Legendre polynomial is much stronger than the second order term (see Figs. \ref{fig:T4} and \ref{fig:T5}), so the second order term can be neglected. The dynamics is dominated by the first order term of the Legendre polynomial. Therefore, we present here only the development of the first order term. It is worth mentioning that numerical simulations were also carried out to verify the magnitude of the solar radiation pressure in first and second order.


\textcolor{black}{Equation (\ref{21}) represents the SRP in the single averaged model considering the Sun in an elliptical and inclined orbit. In \cite{Gkolias,Colombo,Gkoliasb,Krivov,Krivovb,Elisa,Maria} the authors also consider the single averaged model, but the eccentricity of the Sun is not taken into account. Thus, making ${e_{\odot}}=0$ in Eq. (\ref{21}) and after algebraic manipulations, we get the solar radiation pressure equation written in the form}

\begin{equation}\label{22}
\begin{array}{l}
R_{SRP({e_{\odot}}=0)}=\frac{3}{2} e {n_{\odot}}^{2}\beta a_{\odot} a \times ( \cos( {{\odot}}) \cos ( g) \sin( h) \sin( \lambda_{\odot}) - \\[1.5ex]
\qquad \cos( i) \sin ( g) \sin ( h) \cos( \lambda) +\\[1.5ex]
\qquad \sin( i) \sin( { i_{\odot}}) \sin( \lambda_{\odot}) \sin ( g) +\\[1.5ex]
\qquad \cos ( g) \cos( h) \cos ( \lambda_{\odot}) +\\[1.5ex]
\qquad \cos( {i_{\odot}} ) \cos ( i ) \sin
( g) \cos ( h ) \sin ( \lambda_{\odot})),
\end{array}
\end{equation}
where $\lambda_{\odot}=g_{\odot}+h_{\odot}+l_{\odot}$. \textcolor {black}{Comparing Eq. (\ref{22}) with the equation of the solar radiation pressure of \cite{Gkolias} given in Appendix A, page 24 (see also \cite{Krivov,Krivovb,Hamilton}), note that we get the same result except for the term ${n_{\odot}}^{2}\beta a_{\odot} a e$. This happens because our approach is different (see section 2.1) from that presented in \cite{Gkolias}. But using Eqs. (\ref{5}) and (\ref{6}) from this paper and the equations (2.5 a, b) and (2.9) from the book of \cite{McInnes}, we get $2 \frac{A}{m} a e P_{SRP}$, which are the same terms that appear in the equation given in Appendix A, page 24, of \cite{Gkolias}. Therefore, Eq. (\ref{21}) is more general, as we consider the eccentricity of the Earth orbit around the Sun. In this way, we obtain an equation that can be used for other celestial bodies where the eccentricity of the disturber is more elliptical.} Here $i_{\odot}=\varepsilon$ of the equation of \cite{Gkolias}.

\subsection{Oblateness of the Earth}

In this section, we present the equation due to the oblateness of the Earth ($J_{2}$) without much detail, because it is well known in the literature. See, for example, \cite{Paulo}. Considering the equatorial plane of the planet as the reference plane, the disturbing potential, due to zonal terms, can be written in the form:

\begin{equation}\label{23}
\begin{array}{l}
U=-\frac{\mu}{r}\sum_{n=2}^{\infty}\left(\frac{R_{p}}{r}\right)^{n}J_{n}P_{n}(\sin \phi ),
\end{array}
\end{equation}
where $\mu$ is the gravitational constant of the planet, $R_{p}$ is the equatorial radius of the planet, $P_{n}$ are the Legendre polynomials, the angle $\phi$ is the latitude of the orbit with respect to the equator of the planet. Using spherical trigonometry we have $\sin \phi =\sin i \sin(f+g)$. The Legendre polynomials for $J_{2}$ can be written in the form

\begin{equation}\label{24}
\begin{array}{l}
P_{2}(\sin\phi)=\frac{1}{2}(3s^{2}\sin^{2}(f+g)-1),
\end{array}
\end{equation}
where $s=\sin i$ and $c=\cos i $. We write the potential given by Eq. (\ref{23}) as a function of the orbital elements. Invoking Eq. (\ref{24}) and the equation $\mu=n^{2}a^{3}$ ($n$ is the mean motion of the satellite), we get

\begin{equation}\label{25}
\begin{array}{l}
U_{20}=-\frac{1}{2}{\frac {{a}^{3}}{{r}^{3}}\epsilon{n}^{2}( 3{s}^{2}(\sin(f+g))^{2}-1)},
\end{array}
\end{equation}
where $\epsilon=J_{2}R_{p}^{2}$. To write the disturbing potential, we apply the single-averaged model. The development of the equations is carried out in closed form to avoid expansions in eccentricity and inclination. For this, it was necessary to perform algebraic manipulations where we used known equations of celestial mechanics, namely equations $a/r=(1+e\cos(f))/(1-e^2)$ and $dl=\frac{1}{\sqrt{1-e^2}}\frac{r^2}{a^2}df$. After performing the single-average over the true anomaly of the spacecraft, using Eq. (\ref{25}), and after some algebraic manipulations, we get

\begin{equation}\label{26}
\begin{array}{l}
<R_{J2}>=-\frac{1}{4}\frac {\epsilon}{{(1-{e}^{2}) ^{3/2}}}{n}^{2}( 3{s}^{2}-2),
\end{array}
\end{equation}
\textcolor{black}{Equation (\ref{26}), is in accordance with \cite{Krivov,Krivovb,Hamilton} and references therein.}

\subsection{Third-body perturbation}

The disturbing potential in elliptical and inclined orbit due to the third body is presented in \cite{Carvalhoa}, considering the single averaged model, where the authors made several tests and comparisons to validate the equation. Including comparisons with the complete model (direct numerical simulation using the Mercury package), in all tests the single averaged analytical model proved to be quite accurate. Considering the cartesian system fixed in the Earth, let us take the reference plane as the equator of the planet. The disturbing function of the motion of the artificial satellite disturbed by a third body is written in the form

\begin{equation}\label{27}
\begin{array}{l}
R_{\odot}=\frac{G(m_{\bigoplus}+ m_{\odot})r^{2}}{2r_{\odot}^{3}}(3cos^{2}(\psi)-1),
\end{array}
\end{equation}
where $m_{\bigoplus}$ is the mass of the central body, $m_{\odot}$ is the mass of the disturbing body, $G$ is the universal gravitational constant, $r$ and $r_{\odot}$ are the position vector of the satellite and of the third body, respectively. Here $\psi$ is the angular distance between the third body and the satellite. As developed in \cite{Carvalhoa}, now replacing Eqs. (\ref{12})-(\ref{16}) in Eq. (\ref{27}), after algebraic manipulations we obtain,

\begin{equation}\label{28}
\begin{array}{l}
R_{2SA}=\frac {15\mu'{n_{\odot}}^{2}{a}^{2}}{8}({{\frac {a_{\odot}}{{r_{\odot}}}}})^{3}\times \\[1.5ex]
\qquad ( {e}^{2}{D}^{2}\cos( 2{f_{\odot}
}+2{g_{\odot}}-2{h_{\odot}}-2g+2h) +\\[1.5ex]
\qquad{e}^{2}{A}^{2}\cos( 2{f_{\odot}}+2{g_{\odot}}-2{h_{\odot}}+2g+2h)+\\[1.5ex]
\qquad {e}^{2}
{C}^{2}\cos( 2{f_{\odot}}+2{g_{\odot}}+2{h_{\odot}}-2g-2h)+\\[1.5ex]
\qquad{e}^{2}{B}^{2}\cos( 2{f_{\odot}}+2{g_{\odot}}+2{h_{\odot}
}+2g-2h) +\\[1.5ex]
\qquad 2{e}^{2}CE\cos( 2{ f_{\odot}}+2{g_{\odot}}+{
h_{\odot}}-2g-h) -\\[1.5ex]
\qquad2{e}^{2}BE\cos( 2{f_{\odot}}+2{g_{\odot}
}+{h_{\odot}}+2g-h) +\\[1.5ex]
\qquad 2{e}^{2}DE\cos( 2{g_{\odot}}-2g+h+
2{f_{\odot}}-{h_{\odot}}) -\\[1.5ex]
\qquad2{e}^{2}AE\cos( 2{g_{\odot}}+2g
+h+2{f_{\odot}}-{h_{\odot}} ) + \\[1.5ex]
\qquad 6/5DA( {e}^{2}+2/3)
\cos ( 2{f_{\odot}}+2{g_{\odot}}-2{h_{\odot}}+2h) +\\[1.5ex]
\qquad6/5BC( {e}^{2}+2/3) \cos( 2{f_{\odot}}+2{g_{\odot}}+2{
h_{\odot}}-2h) +\\[1.5ex]
\qquad 6/5E( B-C)( {e}^{2}+ 2/3) \cos( 2{f_{\odot}}+2{g_{\odot}}+{h_{\odot}}-h) + \\[1.5ex]
\qquad 6/5E( A-D)( {e}^{2}+2/3) \cos( 2{g_{\odot}}+h+2{f_{\odot}}-{h_{\odot}}) +\\[1.5ex]
\qquad{e}^{2} ( 2CD+{E}^{2}) \cos( -2g+2{f_{\odot}}+2{g_{\odot}}) +\\[1.5ex]
\qquad {e}^{2}( 2AB+{E}^{2}) \cos( 2g+2{f_{\odot}}+2{ g_{\odot}}
) +\\[1.5ex]
\qquad 2{e}^{2}AC\cos( 2g+2h-2{h_{\odot}})+\\[1.5ex]
\qquad 2{e}^{2}BD\cos( 2{h_{\odot}}+2g-2h) +\\[1.5ex]
\qquad 2{e}^{2}E( B
-D) \cos( 2g-h+{h_{\odot}}) +\\[1.5ex]
\qquad 2{e}^{2}E( A-C) \cos ( 2g+h-{h_{\odot}}) +\\[1.5ex]
\qquad 6/5( AC+BD-{E}^{
2})( {e}^{2}+2/3) \cos( 2{f_{\odot}}+2{g_{\odot}}) +\\[1.5ex]
\qquad 6/5( AB+CD)( {e}^{2}+2/3) \cos( -2{h_{\odot}}+2h) -\\[1.5ex]
\qquad 6/5E( A+B-C-D
)( {e}^{2}+2/3) \cos ( h-{h_{\odot}}) +\\[1.5ex]
\qquad 2{e}^{2}( AD+BC-{E}^{2}) \cos ( 2g) +\\[1.5ex]
\qquad 3/5
( {A}^{2}+{B}^{2}+{C}^{2}+{D}^{2}+2{E}^{2}-2/3 )
( {e}^{2}+2/3)).
\end{array}
\end{equation}
Here we use the relation $G(m_{\bigoplus}+m_{\odot})=n_{\odot}^{2}a_{\odot}^{3}=\mu_{\odot}$, where $n_{\odot}$ is the mean motion of the disturbing body and $\mu'=\frac{m_{\odot}}{(m_{\bigoplus}+m_{\odot})}$. Now, for the mean anomaly of the third body ($l_{\odot}$) to appear explicitly in Eq. (\ref{28}), we use the following expansions given by Eqs. (\ref{19}), (\ref{20}) and (\ref{143}) (see \cite{Murray}). The final equation for the single-averaged disturbing potential is given in Appendix A and represents the disturbing body \textcolor{black} {(Sun and Moon)} in elliptical and inclined orbit ($R_{2SA}$). \textcolor{black}{In \cite{Yokoyama2} the single averaged potential for the solar part is given by equation (5). In this equation, in the sum of quadratic terms $A^{2}+B^{2}+...$, the term $D^{2}$ is missing. On the other hand, in the term: $-3EZ(A-C)\cos(2g+\omega-\omega_{\odot})$ there are two errors: the first negative signal should be changed to positive and the argument $g$ inside the cosine should be changed to $w$. The following terms: $+3AEZ\cos(2w+2f_{\odot}+2w_{\odot}+\Omega-\Omega_{\odot})$, $-3EZ(B-D)\cos(2w-\Omega+\Omega_{\odot})$, $+3BEZ\cos(2w+2f_{\odot}+2w_{\odot}-\Omega+\Omega_{\odot})$, must be multiplied by (-1). Fixed typos in equation (5) of \cite{Yokoyama2}, we get the same results for the single averaged system. See also \cite{Colombof} and references therein.}

\begin{equation}\label{143}
\begin{array}{l}
{\left( {\frac {a_{\odot}}{{\it r_{\odot}}}}\right)}^{3}=1+3\,e_{\odot}\cos \left( {\it l_{\odot}} \right) +3/2
\,{e_{\odot}}^{2} \left( 1+3\,\cos \left( 2\,{\it l_{\odot}} \right)  \right).
\end{array}
\end{equation}

Finally, the disturbing potential is written in the form

\begin{equation}\label{144}
\begin{array}{l}
R=<R_{SRP}>+<R_{J2}>+<R2SA_{Sun}+R2SA_{Moon}>.
\end{array}
\end{equation}

Replacing Eq. (\ref{144}) in the Lagrange planetary equations and integrating numerically, we obtain the results presented in the next section.

\section{Results}

\begin{figure}[!htb]
\centering
\subfigure[]
{
\includegraphics[angle=0,width=0.45\linewidth]{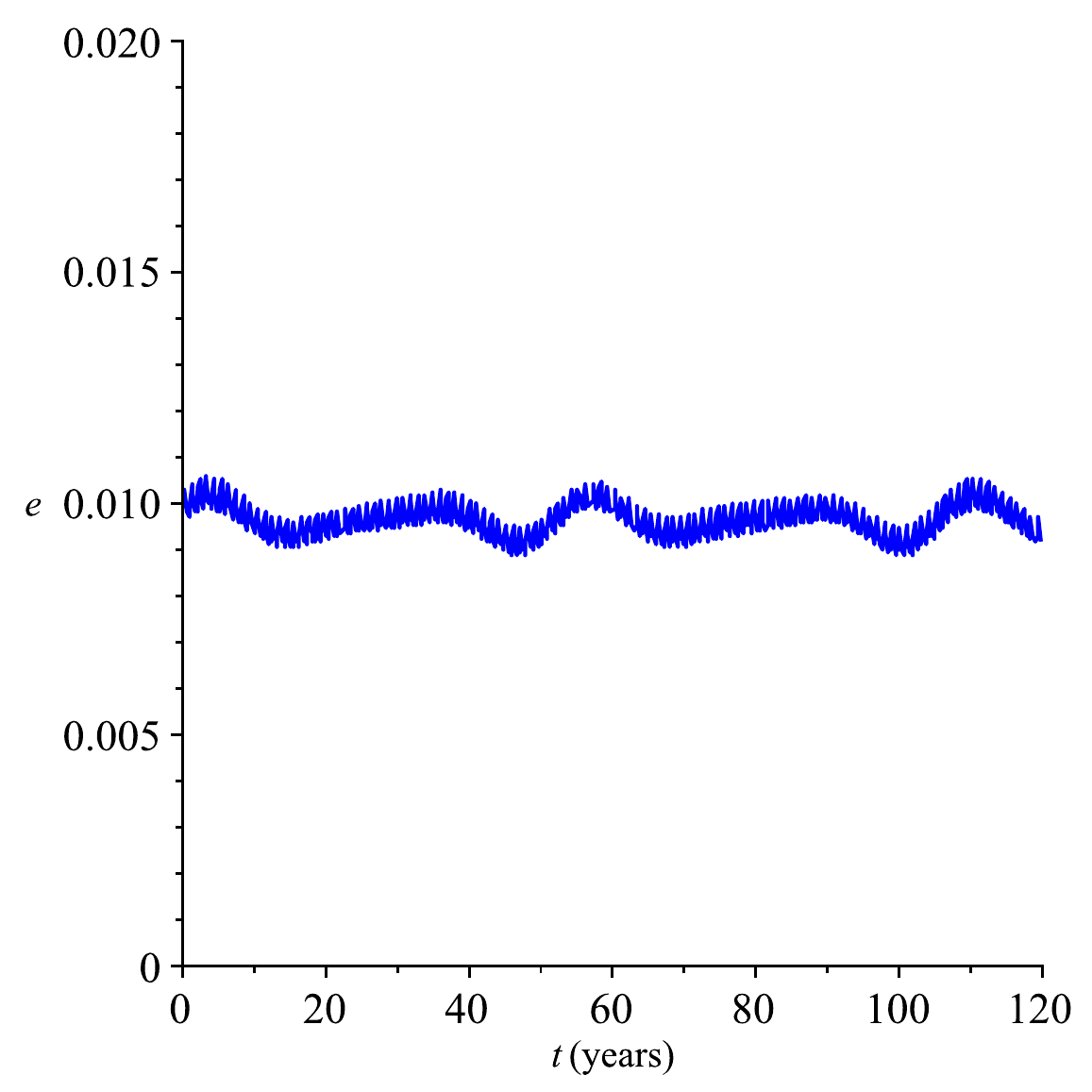}
\label{fig:Fig1}
}
\subfigure[]
{
\includegraphics[angle=0,width=0.45\linewidth]{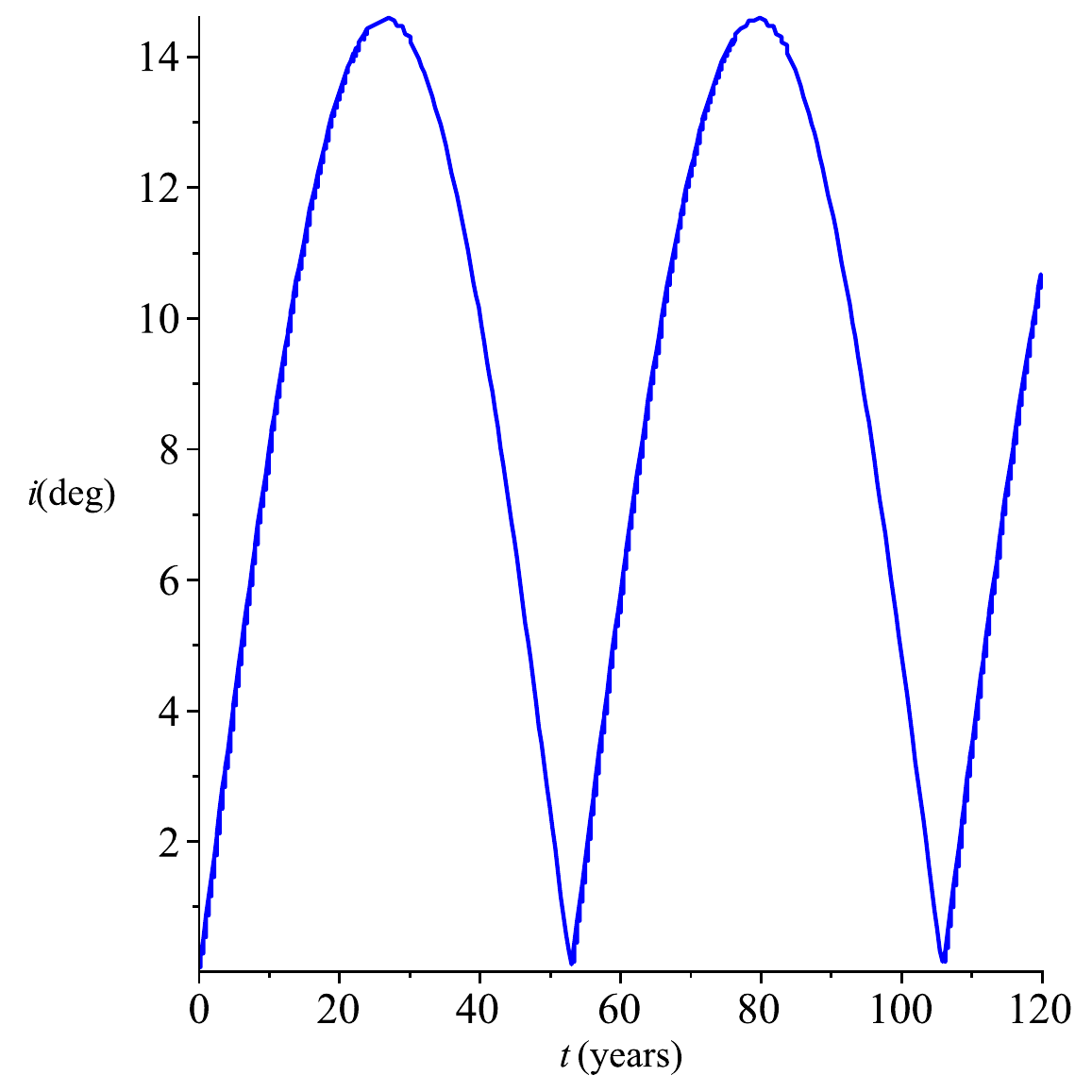}
\label{fig:Fig2}
}
\\[3.5ex]
\qquad
\caption{Initial conditions given by \cite{Gkolias}: $a=42165$ km, $e=0.01$, $i=0.1^{\circ}$, $g=50^{\circ}$, $h=10^{\circ}$ and $A/m=0.012m^{2}/kg$. Disturbing potential: $R_{J2}+R_{SRP}+R2SA_{Sun}+R2SA_{Moon}$. (a) $e$ $\times$ $t$ (b) $i$ $\times$ $t$.}
\end{figure}

To check the equations, we use the initial conditions given in \cite{Gkolias} to compare the results using our disturbing potential, given by Eq. (\ref{144}). Equation (\ref{144}) is replaced in the Lagrange planetary equations and integrated numerically via Maple software. Figures \ref{fig:Fig1} and \ref{fig:Fig2} show the behavior of the eccentricity and inclination, considering the value of the area-to-mass ($A/m$) ratio of a usual satellite ($A/m=0.012$ $m^2/kg$). The result is in agreement with \cite{Gkolias}. Note that, in \cite{Gkolias}, the authors considered in the force model the contributions of the Earth geopotential, perturbation of the third body of the Sun and Moon, the Earth general precession, and the effect of the solar radiation pressure. For the geopotential model, a fourth order and degree truncation is chosen. For zonal harmonics, the first-order averaged contributions were considered and the second order contribution due to $J_{2}$ ($J_{2}^{2}$) was also included. As for the tesseral effects, only the resonant contributions relevant to the geosynchronous orbits were considered. Although the model considered in our work is more simplified, from the point of view of the perturbations considered, we obtain basically the same figures presented in \cite{Gkolias} (see its figure 1). \textcolor{black}{We performed some simulations considering the harmonics $J_{2}$ up to the $J_{6}$ and $C_{22}$, but the SRP effect is predominant in our investigation (considering high mass-area values) and, thus, did not change the results presented in this work. Therefore, we only consider the $J_{2}$ term.}

\begin{figure}[!htb]
\centering
\subfigure[]
{
\includegraphics[angle=0,width=0.45\linewidth]{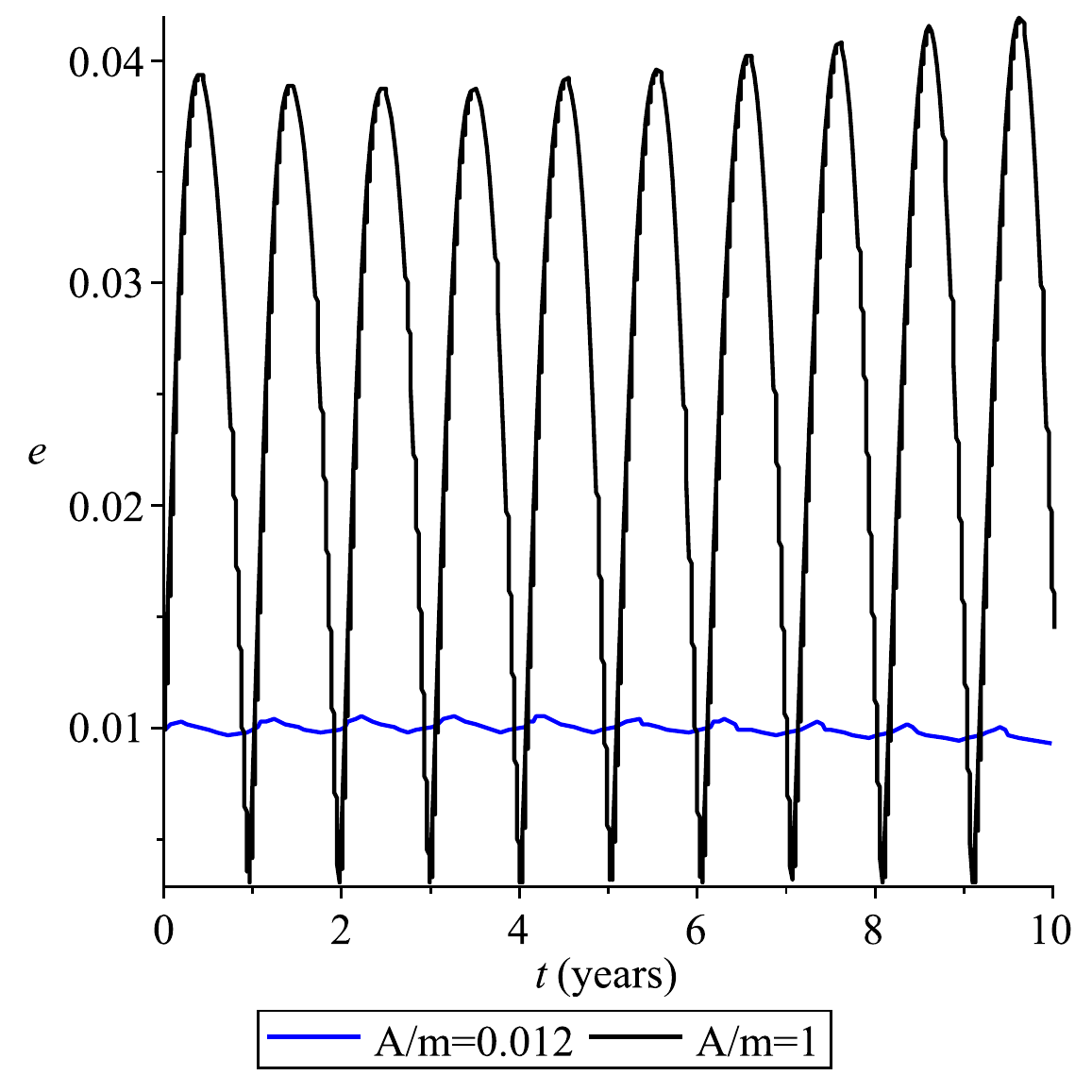}
\label{fig:Fig3}
}
\subfigure[]
{
\includegraphics[angle=0,width=0.45\linewidth]{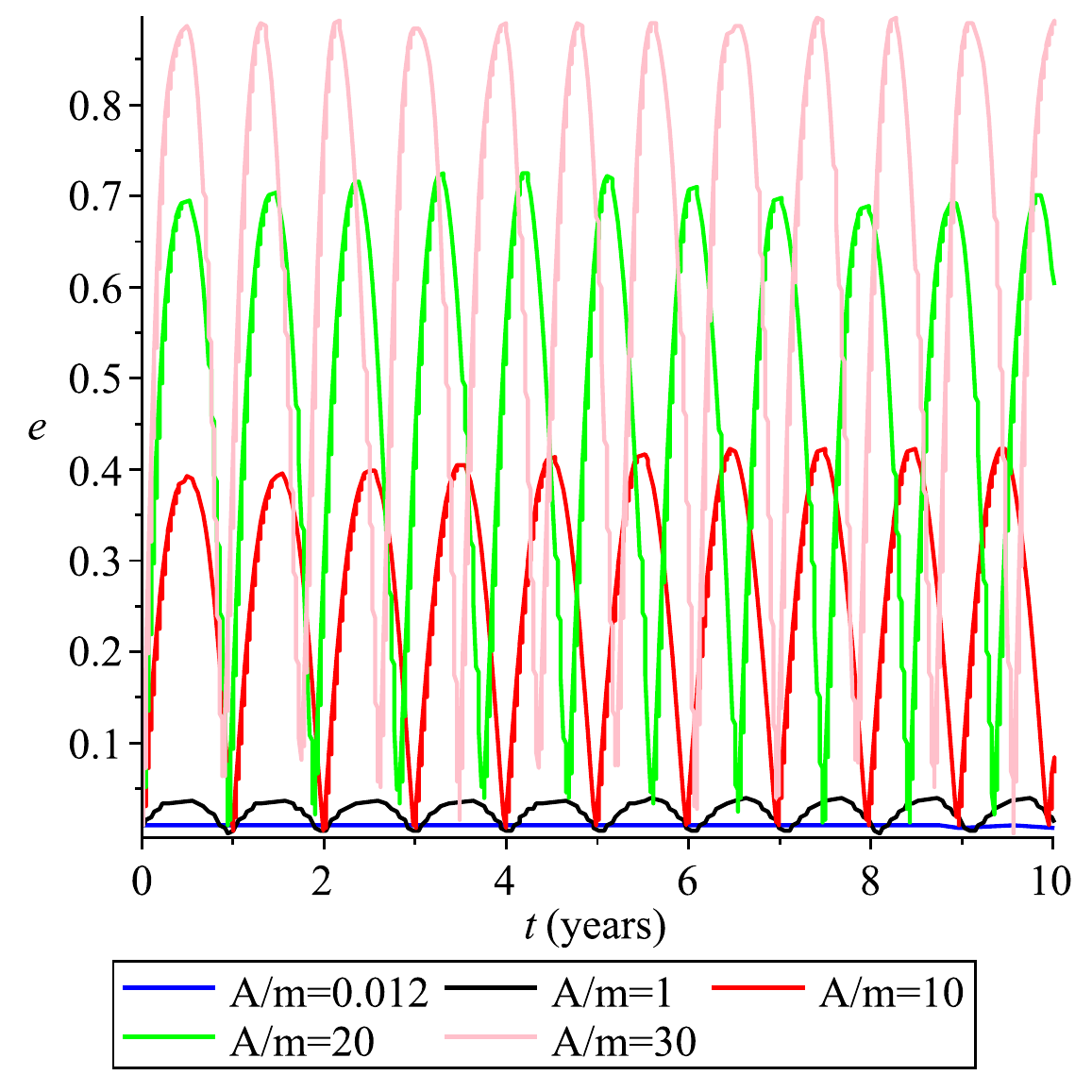}
\label{fig:Fig4}
}
\\[3.5ex]
\qquad
\caption{Initial conditions given by \cite{Gkolias}: $a=42165$ km, $e=0.01$, $i=0.1^{\circ}$, $g=50^{\circ}$, $h=10^{\circ}$ and $A/m=0.012m^{2}/kg$. Disturbing potential: $R_{J2}+R_{SRP}+R2SA_{Sun}+R2SA_{Moon}$. (a) Two $A/m$ rate values (b) Various $A/m$ rate values.}
\end{figure}

As we can see in Fig. \ref{fig:Fig1}, a debris can remain in GEO orbit for hundreds of years, as the eccentricity varies very little over time. Now, increasing the value of the area-to-mass ratio to $1 m^{2}/kg$ (see Fig. \ref{fig:Fig3}), we observe that the eccentricity presents a considerable growth when compared to Fig. \ref{fig:Fig1}. Thus, Fig. \ref{fig:Fig4} shows the characteristic of the eccentricity of the debris for different values of the area-mass ratio. Note that the increase in eccentricity is proportional to the area-mass ratio and that it grows very fast. \textcolor{black}{It is important to emphasize that the increase in eccentricity has already been observed in many previous works, such as \cite{Lucking,Luckingb,Colombob,Hamilton,Krivov,Krivovb}.}

\begin{figure}
\centering
\resizebox{0.55\columnwidth}{!}{\includegraphics{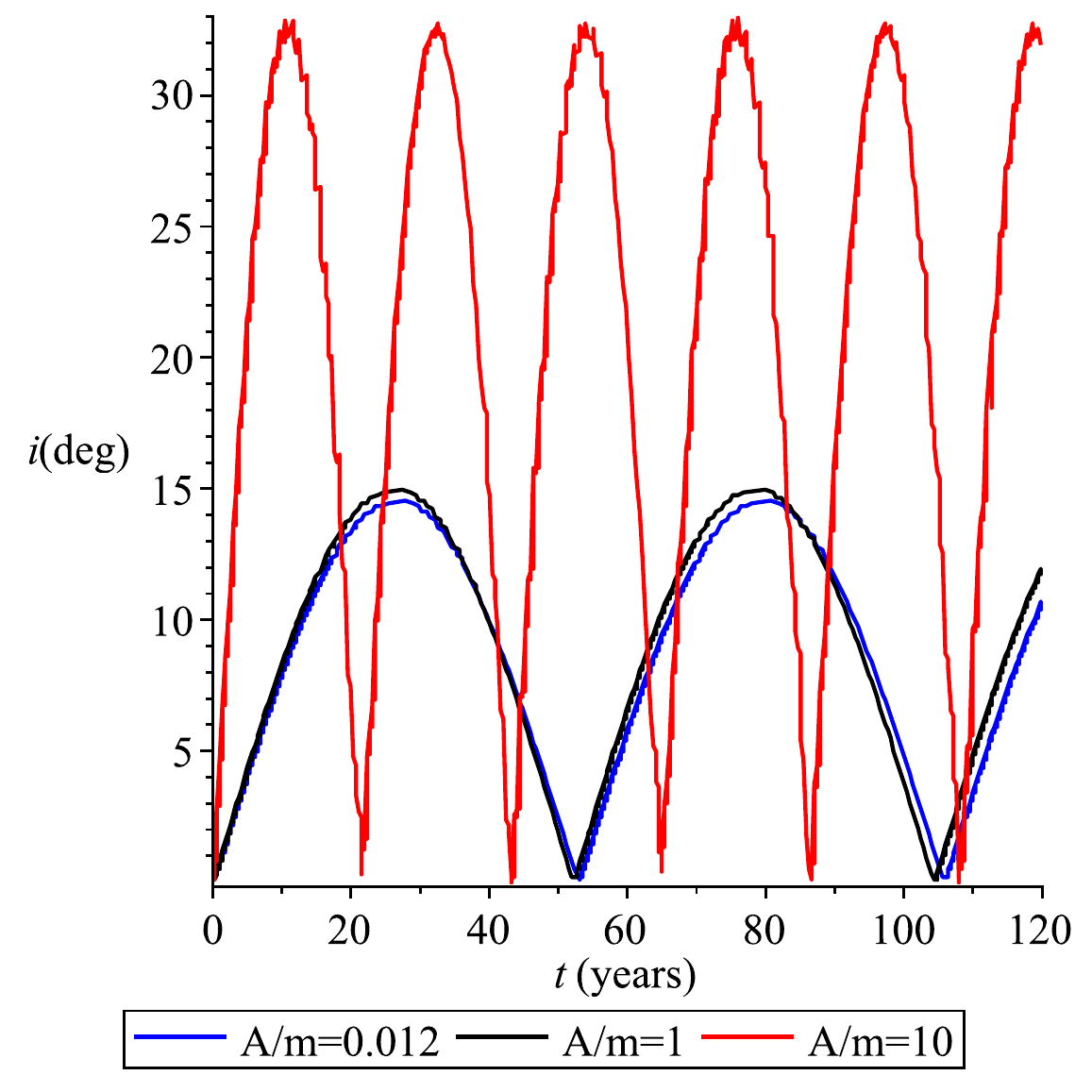}}
\caption{Initial conditions given by \cite{Gkolias}: $a=42165$ km, $e=0.01$, $i=0.1^{\circ}$, $g=50^{\circ}$, $h=10^{\circ}$ and $A/m$ in $ m^{2}/kg$. Disturbing potential: $R_{J2}+R_{SRP}+R2SA_{Sun}+R2SA_{Moon}$.}
\label{fig:Fig5}       
\end{figure}

Figure \ref{fig:Fig5} shows that the inclination is also strongly disturbed with the increase in the area-mass ratio, which can contribute, for example, to deviate the debris from certain collision routes. However, the inclination needs a larger time to reach great amplitude.

\begin{figure}
\centering
\includegraphics[scale=0.45]{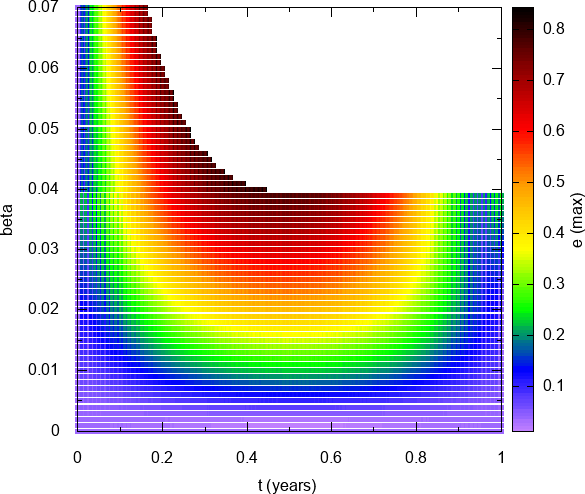}
\caption{E06321D debris. Initial conditions: $a=41400$ km, $e=0.035$, $i=7^{\circ}$, $g=0^{\circ}$, $h=0^{\circ}$. Disturbing potential: $R_{J2}+R_{SRP}+R2SA_{Sun}+R2SA_{Moon}$.}
\label{fig:RR}
\end{figure}

\begin{figure}
\centering
\includegraphics[scale=0.45]{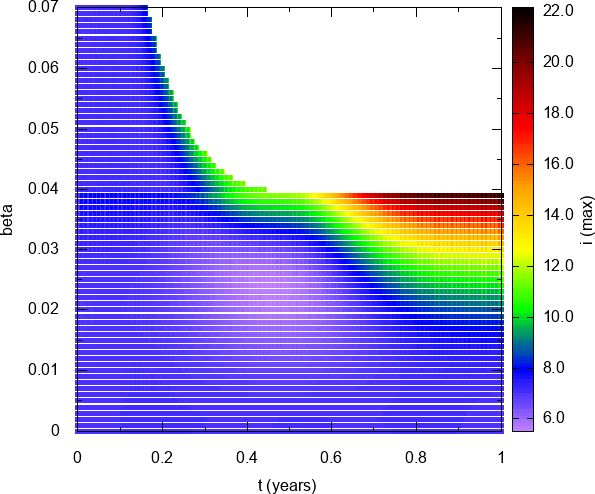}
\caption{E06321D debris. Initial conditions: $a=41400$ km, $e=0.035$, $i=7^{\circ}$, $g=0^{\circ}$, $h=0^{\circ}$. Disturbing potential: $R_{J2}+R_{SRP}+R2SA_{Sun}+R2SA_{Moon}$.}
\label{fig:RRR}
\end{figure}

\begin{figure}[!htb]
\centering
\subfigure[]
{
\includegraphics[angle=0,width=0.45\linewidth]{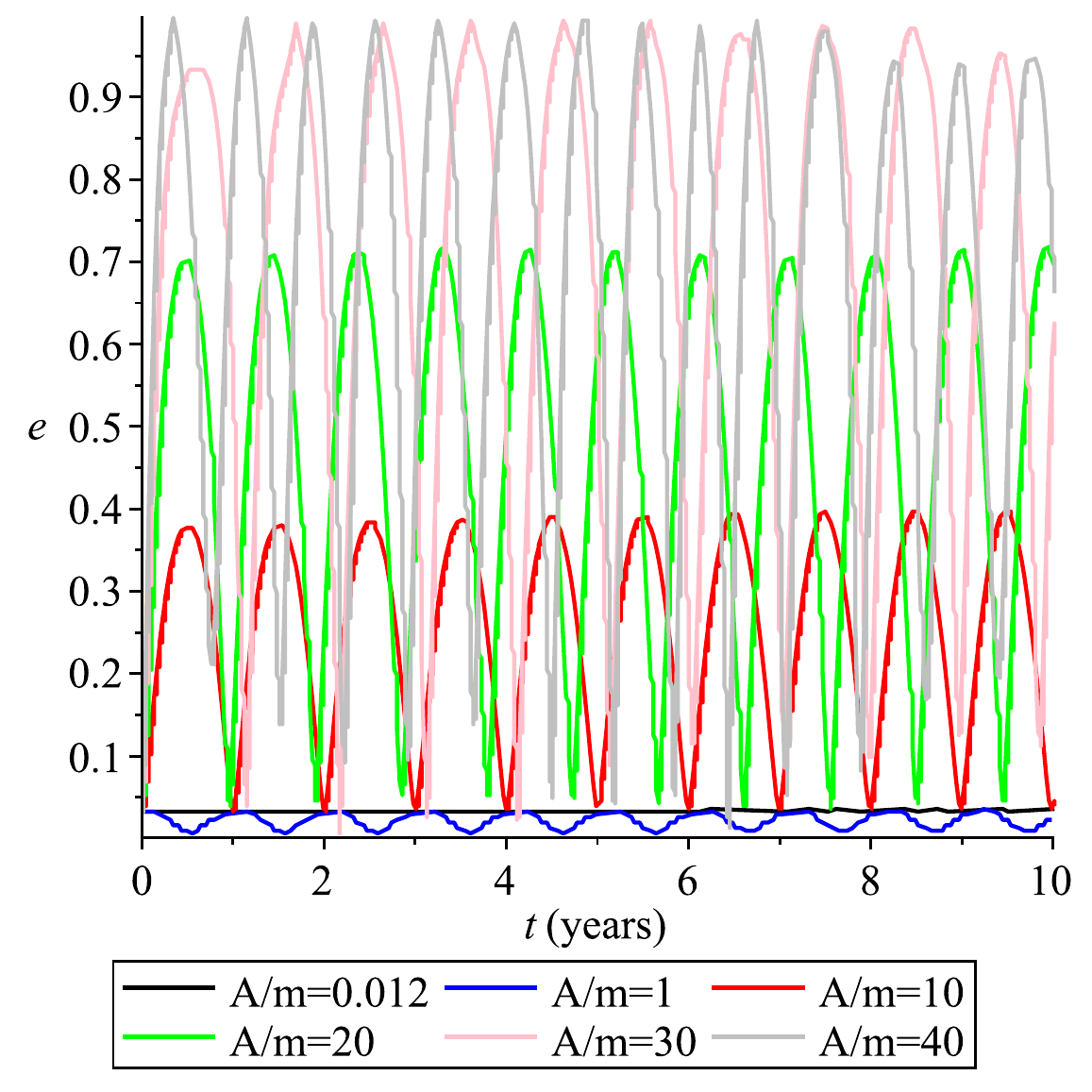}
\label{fig:W1}
}
\subfigure[]
{
\includegraphics[angle=0,width=0.45\linewidth]{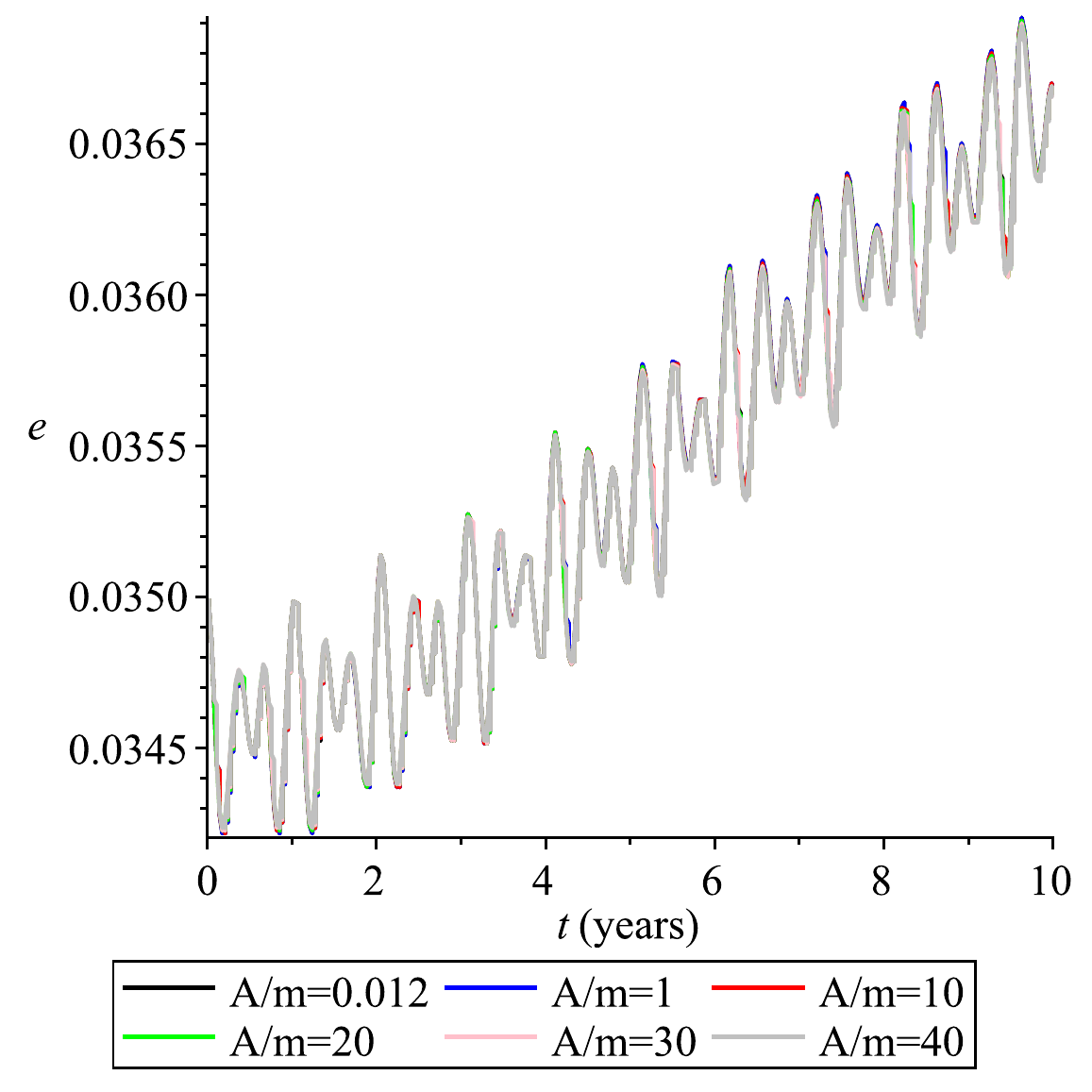}
\label{fig:W4}
}
\\[3.5ex]
\qquad
\caption{E06321D debris. Initial conditions: $a=41400$ km, $e=0.035$, $i=7^{\circ}$, $g=0^{\circ}$, $h=0^{\circ}$ and $A/m$ in $ m^{2}/kg$. (a) $e$ $\times$ $t$. Disturbing potential: $R_{J2}+R_{SRP}+R2SA_{Sun}+R2SA_{Moon}$. (b) $e$ $\times$ $t$. Disturbing potential: $R_{J2}+R2SA_{Sun}+R2SA_{Moon}$.}
\end{figure}

\begin{figure}[!htb]
\centering
\subfigure[]
{
\includegraphics[angle=0,width=0.45\linewidth]{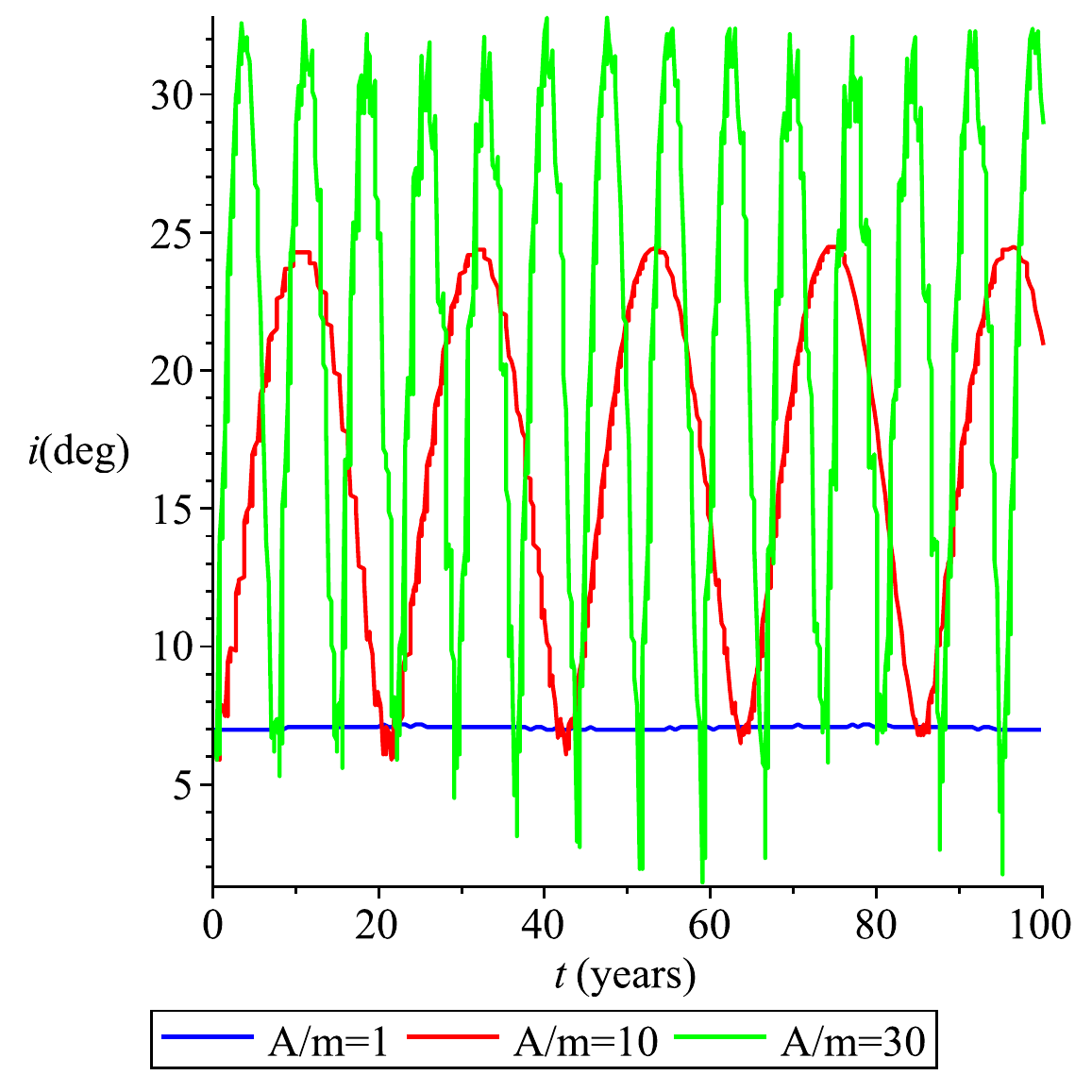}
\label{fig:W2}
}
\subfigure[]
{
\includegraphics[angle=0,width=0.45\linewidth]{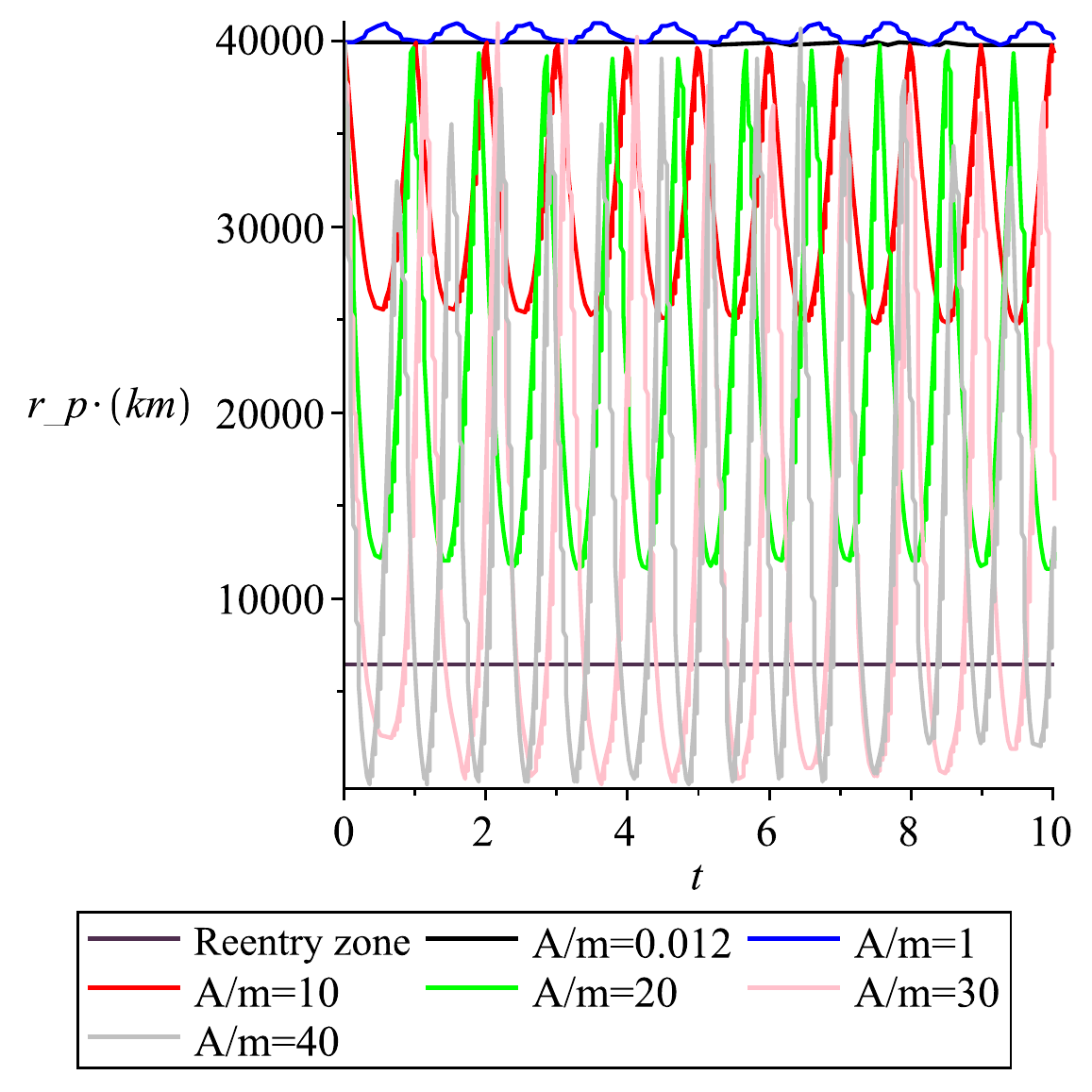}
\label{fig:W3}
}
\\[3.5ex]
\qquad
\caption{E06321D debris. Initial conditions: $a=41400$ km, $e=0.035$, $i=7^{\circ}$, $g=0^{\circ}$, $h=0^{\circ}$ and $A/m$ in $ m^{2}/kg$. Disturbing potential: $R_{J2}+R_{SRP}+R2SA_{Sun}+R2SA_{Moon}$. (a) $i$ $\times$ $t$ (b) $r_p=a(1-e)$ $\times$ $t$.}
\end{figure}

The orbital data of the debris that are located in the geostationary orbit used in this work are obtained from the site "stuffin.space" (http://stuffin.space/
). This platform is updated daily with orbit data from "Space-Track.org". As in \cite{Gkolias}, here we also assume that the re-entry condition of 120 km above the surface of the Earth ($a_{re-entry}=R_{Earth} + 120$ km) for a satellite at the GEO region, the re-entry value for the eccentricity is $e_{re-entry}\approx 0.846$. The data of the space debris used in the simulations shown in Figs. \ref{fig:RR}, \ref{fig:RRR}, \ref{fig:W1}, \ref{fig:W4}, \ref{fig:W2} and \ref{fig:W3}, is obtained from the site "stuffin.space", which is called E06321D debris. Figure \ref{fig:RR} shows a color map where we vary the value of the $\beta$ parameter to analyze the corresponding values of the area-mass ratio so that the debris can approach the surface of the Earth. \textcolor{black}{We consider a dynamic analysis taking into account high values of the $A/m$ ratio (see Figure \ref{fig:RR}), but, due to the limitation of the solar sail reaching these very high values, our study is academic and do not a give definitive solution
for space debris mitigation. But, as the results show, it is possible to do the removal for the $A/m$ values presented in the figures. In the continuation of this work, we will investigate, through dynamic maps, the initial conditions that contribute to find lower values of the $A/m$ parameter.} The eccentricity growth is already visible for the parameter values around approximately $\beta=0.03$, which corresponds to a value of $A/m$ around 20 $m^{2}/kg$ (to make this calculation use Eq. (\ref{5})). For values larger than or equal to 0.04 (30 $m^{2}/kg$) the debris reaches the region of the semimajor axis that we define as the reentry zone ($R_{Earth}$ + 120 km) in a short time. Figure \ref{fig:RR} shows that, for $\beta$ values larger than 0.04, the debris are ejected, as they acquire high eccentricities, as shown by the color scale. This ensures the removal of the debris from the GEO zone very quickly. Figure \ref{fig:RRR} shows a color map equivalent to Fig. \ref{fig:RR}, but now the color scale shows the maximum inclination values. Unlike the previous case, a larger time is necessary to achieve a great variation in the inclination. In \textcolor{black}{\cite{Miguel} (and references therein) the size and shape of the solar sail is discussed according to the available technology. The size of the sail panels is a critical physical parameter for predicting deorbiting times, for a fixed mass of the spacecraft. The larger the cross area is, the faster the deorbiting. The authors considered three different constructible structures according to the guidelines of \cite{Dalla}. In \cite{Dalla} the authors provide a method to assess the constructability of a square solar sail for certain values of the payload mass and the area-to-mass ratio, with current technology restrictions. The sail module mass with the sail sidelength is given by two best fit equations, depending on the chosen technology. That is, there is a technological limitation for large values of the $A/m$ ratio, thus, our results show the dynamics of an idealized solar sail. But, with a refinement of the model and the inclusion of other perturbations, it is possible to obtain results with the lowest value of the $A/m$ parameter, which will be the object of future work.}

In Fig. \ref{fig:W1}, when we increase the $A/m$ ratio of the debris, the eccentricity grows very quickly and in less than a year, we obtain extreme values for the eccentricity. In the case of a space vehicle or debris in the GEO region, it is known that the eccentricity is very little disturbed (see the case where $A/m=0.012$ $m^{2}/kg$), in such a way that the vehicle remains in orbit for hundreds of years. Thus, attaching a solar sail, or even opening a solar sail at the end of the lifetime of the artificial satellite, which is already installed in its compartment, we can remove the debris from the GEO zone very quickly. \textcolor{black}{Note that in \cite{Gkolias} the $A/m$ values of 0.0012 $m^{2}/kg$ and 1 $m^{2}/kg$ were chosen according to the available technology. Thus, the values presented in this paper are not technologically possible. Therefore, due to the limits of achieving this in practice, our results only show the dynamics of an idealized solar sail.} To compare the evolution of the eccentricity without considering the solar radiation pressure, we present Fig. \ref{fig:W4}, where this perturbation is not taken into account. Note that, in this case, the eccentricity of the debris vary very little over time for any value of the area-mass ratio considered (see figure legend). Now, comparing Figs. \ref{fig:W1} and \ref{fig:W4}, we can emphasize how the SRP influences the dynamics of the debris over time, increasing the eccentricity very quickly. In the case of the inclination, this growth is slower, but in any case there is a certain variation in the inclination as we increase the value of the $A/m$ ratio, see Fig. \ref{fig:W2}. \textcolor{black}{In this case, the solar sail can be used to modify the position of the debris, for example to change the orbit parameters and switch to a “safer” orbit}. Finally, Fig. \ref{fig:W3} shows the position of the perigee considering different values of the $A/m$ ratio. Note that, for the $A/m$ values equal to 30 and 40 $m^{2}/kg$, the debris, besides reaching the reentry zone, collides with the Earth surface. The area-to-mass ratios between 20 and 30 $m^{2}/kg$ are the values for the solar sail to take the debris (with the SRP as a proponent mechanism) for the so-called reentry zone defined above, where $a_{re-entry}=R_{Earth} + 120$ km. It is worth noting that, for lower values of the area-mass ratio they also increase the eccentricity considerably (see Figs. \ref{fig:W1} and \ref{fig:W3}), but the debris does not reach the reentry zone.

\begin{figure}
\centering
\includegraphics[scale=0.45]{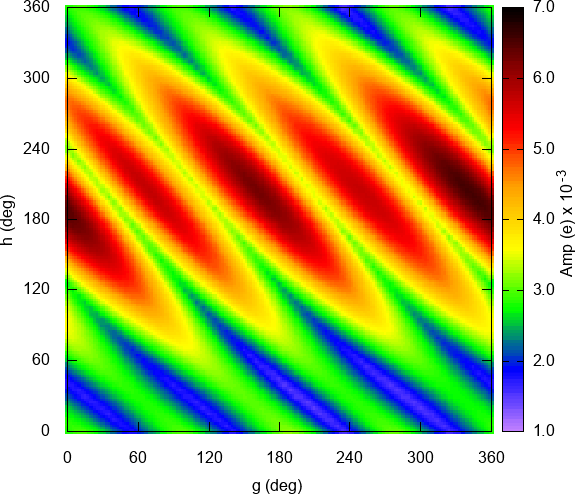}
\caption{E06321D debris. Initial conditions: $a=41400$ km, $e=0.035$, $i=7^{\circ}$, $g=0..360^{\circ}$ and $\Delta g=5^{\circ}$, $h=0..360^{\circ}$ and $\Delta h=5^{\circ}$. Disturbing potential: $R_{J2}+R_{SRP}+R2SA_{Sun}+R2SA_{Moon}$. Integration time: 10 years, $Amp(e)=e_{max}-e_{min}$, $A/m=0.012$ $m^{2}/kg$.}
\label{fig:T1}
\end{figure}

\begin{figure}
\centering
\includegraphics[scale=0.45]{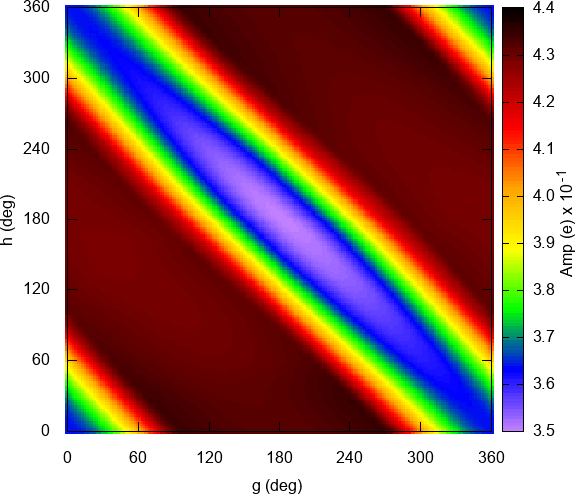}
\caption{E06321D debris. Initial conditions: $a=41400$ km, $e=0.035$, $i=7^{\circ}$, $g=0..360^{\circ}$ and $\Delta g=5^{\circ}$, $h=0..360^{\circ}$ and $\Delta h=5^{\circ}$. Disturbing potential: $R_{J2}+R_{SRP}+R2SA_{Sun}+R2SA_{Moon}$. Integration time: 10 years, $Amp(e)=e_{max}-e_{min}$, $A/m=10$ $m^{2}/kg$.}
\label{fig:T2}
\end{figure}

\begin{figure}
\centering
\includegraphics[scale=0.45]{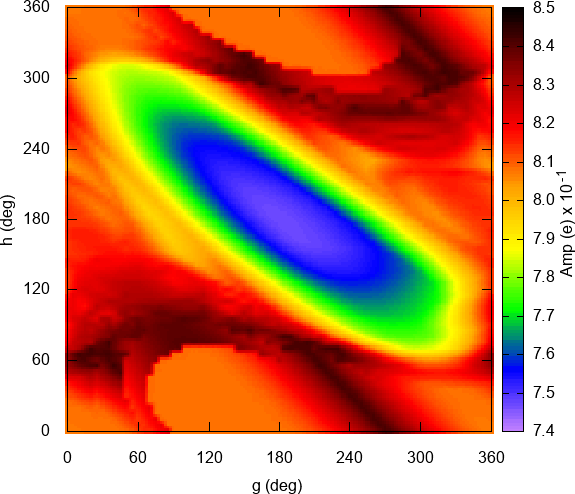}
\caption{06321D debris. Initial conditions: $a=41400$ km, $e=0.035$, $i=7^{\circ}$, $g=0..360^{\circ}$ and $\Delta g=5^{\circ}$, $h=0..360^{\circ}$ and $\Delta h=5^{\circ}$. Disturbing potential: $R_{J2}+R_{SRP}+R2SA_{Sun}+R2SA_{Moon}$. Integration time: 10 years, $Amp(e)=e_{max}-e_{min}$, $A/m=25$ $m^{2}/kg$.}
\label{fig:T3}
\end{figure}

\begin{figure}
\centering
\includegraphics[scale=0.45]{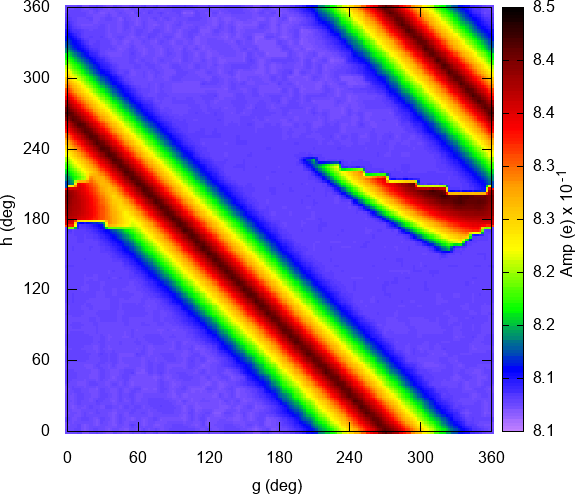}
\caption{06321D debris. Initial conditions: $a=41400$ km, $e=0.035$, $i=7^{\circ}$, $g=0..360^{\circ}$ and $\Delta g=5^{\circ}$, $h=0..360^{\circ}$ and $\Delta h=5^{\circ}$. Disturbing potential: $R_{J2}+R_{SRP}+R2SA_{Sun}+R2SA_{Moon}$. Integration time: 10 years, $Amp(e)=e_{max}-e_{min}$, $A/m=30$ $m^{2}/kg$.}
\label{fig:T6}
\end{figure}

The site "stuffn.space" does not provide data on the perigee argument and the longitude of the ascending node, so Figs. \ref{fig:T1}, \ref{fig:T2}, \ref{fig:T3} and \ref{fig:T6} show a color map where we vary these orbital elements from 0 to 360 degrees (see figure legend) to verify the contribution in the orbit of the debris. \textcolor{black}{The construction of the color maps presented in Figs. \ref{fig:T1}, \ref{fig:T2}, \ref{fig:T3} and \ref{fig:T6} was motivated by the recent publication given by \cite{Gkolias}, where the authors present a complete dynamic mapping taking into account all relevant perturbations. The results are presented in the form of selected stability maps to highlight the underlying mechanisms and their interaction, which can lead to stable graveyard orbits or fast reentry pathways. Although the dynamic system considered in \cite{Gkolias} is more general than the one considered in this work, we obtain some results that can contribute to the studies of space debris mitigation.} The color scale represents the amplitude difference between the maximum and minimum eccentricity values $(Amp(e)=e_{max}-e_{min}$), considering different initial values of the perigee argument and of the longitude of the ascending node. Of course, for higher values of the $A/m$ ratio, the increase in eccentricity is larger, as previously mentioned, but now we find some regions where the ($g, h$) plane contributes to an increase in the eccentricity, as shown in the Figs. \ref{fig:T1}, \ref{fig:T2}, \ref{fig:T3} and \ref{fig:T6}. Note that some regions highlighted in stronger colors show an excitation of the eccentricity of the space debris. Therefore, a convenient choice of ($g, h$) can contribute to the reentry of the debris. It is enough to direct them to these regions, so that the natural perturbations allow the reentry of the debris more quickly. In the case of Fig. \ref{fig:T6}, as the value of the area-mass ratio is high ($A/m=30$ $m^{2}/kg$), the debris, in general, accesses the reentry zone ($a_{re-entry}=R_{Earth} + 120$ km) very quickly, before one year. Even so, it is still possible to choose the ($g, h$) plane to obtain a more accelerated reentry. \textcolor{black}{In \cite{Musci}, the authors show the impact of the $A/m$ ratio on the lifetime of an object in geostationary transfer orbits. In this case, the debris has a high eccentricity and a semi-major axis around 25000 km. That is, increasing $A/m$ reduces, in most cases, the orbital lifetime of the debris. The authors also draw attention to the fact that, depending on the initial condition of the right ascension of the ascending node and the argument of perigee, luni-solar perturbations may have a stronger impact on the orbital lifetime. Here we corroborate that these orbital elements also contribute to the determination of the orbital lifetime of the debris, in particular due to the SRP. In \cite{Elisa,Maria}, the authors show the location of the equilibrium points associated with each resonance effect separately, as a function of ($a, e, i$) for three values of area to mass ratio. Color maps were used to show the effects of the perturbations on eccentricity amplitudes.}

\begin{figure}
\centering
\includegraphics[scale=0.45]{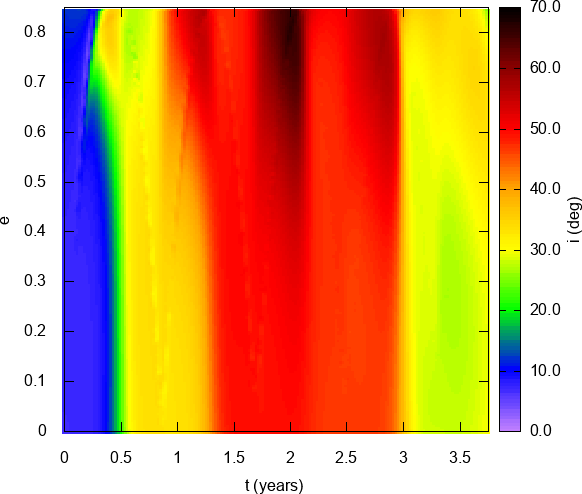}
\caption{E06321D debris. Initial conditions: $a=41400$ km, $e=0.035$, $i=7^{\circ}$, $g=0..360^{\circ}$ and $\Delta g=5^{\circ}$, $h=0..360^{\circ}$ and $\Delta h=5^{\circ}$. Disturbing potential: $R_{J2}+R_{SRP}+R2SA_{Sun}+R2SA_{Moon}$. Integration time: 10 years. Were $A/m=30$ $m^{2}/kg$.}
\label{fig:T7}
\end{figure}

\begin{figure}
\centering
\includegraphics[scale=0.45]{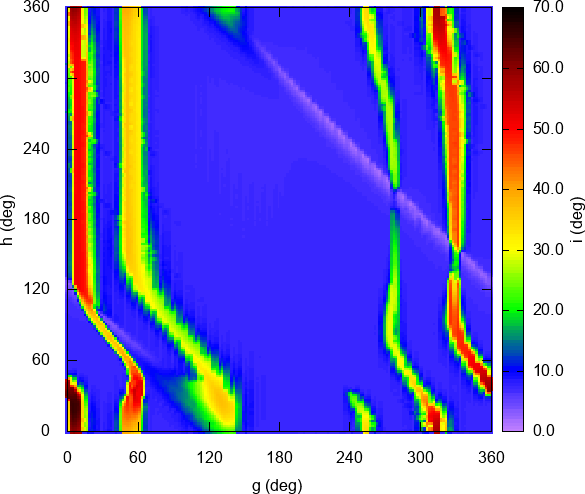}
\caption{E06321D debris. Initial conditions: $a=41400$ km, $e=0.035$, $i=7^{\circ}$, $g=0..360^{\circ}$ and $\Delta g=5^{\circ}$, $h=0..360^{\circ}$ and $\Delta h=5^{\circ}$. Disturbing potential: $R_{J2}+R_{SRP}+R2SA_{Sun}+R2SA_{Moon}$. Integration time: 10 years. Were $A/m=30$ $m^{2}/kg$.}
\label{fig:T8}
\end{figure}

Taking advantage of the simulation data shown in Fig. \ref{fig:T6}, we plotted Fig. \ref{fig:T7}, to show the behavior of the inclination when the eccentricity is increased. The choice of the data in Fig. \ref{fig:T6} is made, in this case because the area-mass ratio is larger and, it makes the eccentricity to be strongly perturbed in a short time. Note that the collision time can be larger for different values of the ($g, h$) plane. For collisions that occur in less than half a year (see Fig. \ref{fig:T7}), the inclination undergoes little variations, that is, the satellite is removed from the GEO region with little inclination, which can cause damage to other vehicles nearby. But, for collisions with a slightly larger time, the inclination increases considerably, reaching the peak in about two years, as shown in Fig. \ref{fig:T7}. Thus, the debris can be removed from the GEO region in a way that they do not cause damage \textcolor{black} {to the} space vehicles present there. Figure \ref{fig:T7} shows that the ($g, h$) plane also contributes to the increase in the inclination and the collision time (or reentry). \textcolor{black}{Finally, Fig. \ref{fig:T8} shows the value of the ($g, h$) plane that contribute to amplify the inclination. Note that the colors that appear in Fig. \ref{fig:T7}, which correspond to the inclination growth, are also represented in Fig. \ref{fig:T8}, but now to show the effect of the perigee argument and the longitude of the ascending node on the orbital inclination.} Therefore, in the way that we carry out this research, allows us to identify the value of the area-mass ratio of the solar sail that should contribute to the removal of the debris more quickly and safely. Information about the values of the perigee argument and of the longitude of the ascending node that we must consider to move the debris and to reach the maximum eccentricity, in a safe time for that the debris is removed from the GEO region with a certain inclination and, therefore, does not cause damage to other space vehicles.

\begin{figure}
\centering
\includegraphics[scale=0.45]{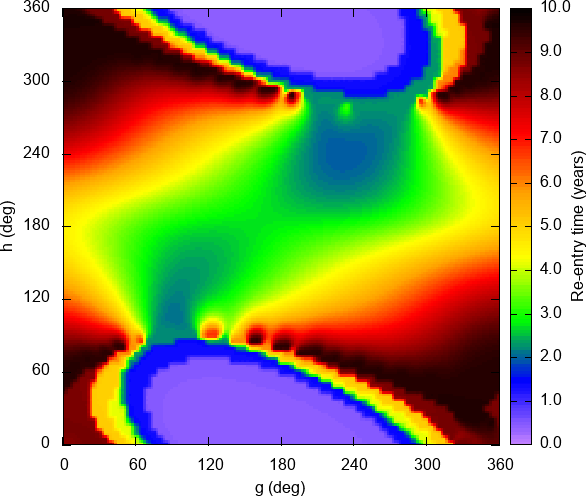}
\caption{E06321D debris. Initial conditions: $a=41400$ km, $e=0.035$, $i=7^{\circ}$, $g=0..360^{\circ}$ and $\Delta g=5^{\circ}$, $h=0..360^{\circ}$ and $\Delta h=5^{\circ}$. Disturbing potential: $R_{J2}+R_{SRP}+R2SA_{Sun}+R2SA_{Moon}$. Integration time: 10 years. Were $A/m=25$ $m^{2}/kg$.}
\label{fig:T9}
\end{figure}

\begin{figure}
\centering
\includegraphics[scale=0.45]{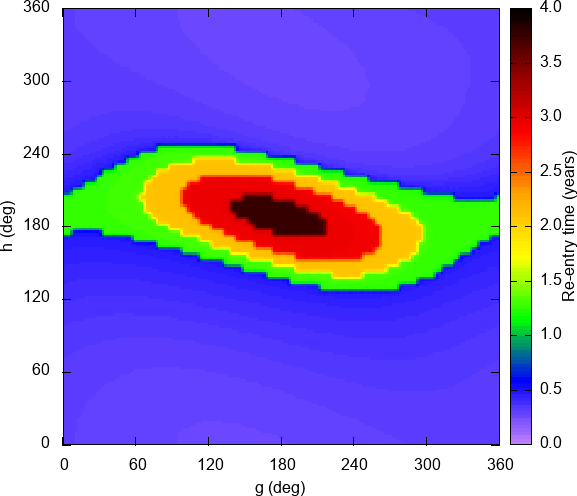}
\caption{E06321D debris. Initial conditions: $a=41400$ km, $e=0.035$, $i=7^{\circ}$, $g=0..360^{\circ}$ and $\Delta g=5^{\circ}$, $h=0..360^{\circ}$ and $\Delta h=5^{\circ}$. Disturbing potential: $R_{J2}+R_{SRP}+R2SA_{Sun}+R2SA_{Moon}$. Integration time: 10 years. Were $A/m=30$ $m^{2}/kg$.}
\label{fig:T10}
\end{figure}

\textcolor{black}{Figure \ref{fig:T9} shows the ($g, h$) plane and collision time (in the color bar) with the $A/m$ value represented in Fig. \ref{fig:T3} ($25 m^{2}/kg$). Note that, in some regions, the re-entry time is very fast and in other regions the debris takes up to 10 years to enter the so-called re-entry zone, as shown in Fig. \ref{fig:T9}. The space debris can be directed to the value of the ($g, h$) plane in such a way that the re-entry time is chosen. Now, Fig. \ref{fig:T10} also shows the collision time in a color scale, as in Fig. \ref{fig:T9}. But, in this case, with $A/m=30$, which is the same value of the area-to-mass ratio shown in Fig. \ref{fig:T6}. Of course, increasing the parameter $A/m$ the eccentricity is more perturbed and thus the collision time is faster than shown in Fig.\ref{fig:T9}. The region of the plane ($g, h$) around ($180 ^{\circ}, 180^{\circ}$) is the one with the longest reentry time. In most values of the plane ($g, h$) the debris is quickly pushed into the re-entry zone.}

\section{Conclusions}

In order to contribute to the research on the \textcolor{black}{Earth-orbit space debris mitigation}, in this work, we consider the main perturbations that act on a solar sail. Namely, the Earth oblateness, the perturbation of the third body (Moon and Sun) and the solar radiation pressure (SRP). \textcolor{black}{The main goal} of this work was to use the solar sail to amplify the growth of the eccentricity and the inclination of space debris in a geostationary orbit. To help clean up the space environment, the Sun is used as a propulsion system, a source of clean and abundant energy, to contribute with the sustainability of the space exploration. For this, we consider a solar sail coupled with the debris, which can be done via specific missions to make the coupling, or for artificial satellites launched with a coupled sail.

We developed the SRP equation based in \cite{Tresaco}, but now using the single averaged model, in which we do an extension to obtain the first and second order terms of the disturbing potential. We show that the first order term dominates the dynamics when compared to the second order term used in \cite{Tresaco}. To perform the simulations, we use orbital data of space debris, which are located in the geostationary orbit, obtained from the site "stuffin.space". From the simulations performed, we found values of the area-mass ratio and the initial conditions of the perigee argument ($g$) and the longitude of the ascending node ($h$) that contribute to the amplification of the eccentricity and inclination of the debris. We show some results using a color map that \textcolor{black} {helps to} determine the value of the ($g, h$) \textcolor{black}{plane} that contribute to the quick and safe removal of space debris, where we find adequate data for the object to be removed with a certain inclination to avoid a collision with others nearby vehicles. In the continuation of this work, the perturbation due to atmospheric drag \textcolor{black}{and the effect of the Earth shadow will be included} to analyze the evolution of space debris during the re-entry into the Earth atmosphere. The debris evolution in LEO and MEO orbits will also be analyzed.

\section{Appendix A}

The disturbing potential due to the third-body \textcolor{blue} {(Sun and Moon)} in an elliptical and inclined orbit, obtained via the single-averaged model, is written in the form (\cite{Carvalhoa})

\begin{align*}\label{34}
R2_{SA}=&{\frac {255 k n_{\odot}^{2}{a}^{2}}{256}} \times \\
	&  ( {\frac {7 e_{\odot}{e}^{2}  ( {
			c_{\odot}}-1  ) ^{2}  ( c-1  ) ^{2}\cos  ( -{  l_{\odot}}+2
		g-2 h-2 \lambda_{\odot}+4 {  h_{\odot}}  ) }{17}}+  \\
	& {\frac {7 e_{\odot}{e}^{2} ( {  c_{\odot}}-1  ) ^{2}  ( 1+c  ) ^{2}\cos  ( {l_{\odot}}+2 g+2 h+2 \lambda_{\odot}-4 {  h_{\odot}}  ) }{17}}-\\
	 & 1/17 e_{\odot}{e}^{2}( 1+{  c_{\odot}}  ) ^{2}  ( c-1  ) ^{2}\cos  ( 2g-2 h+2 \lambda_{\odot}-{  l_{\odot}}  ) -\\
	 & 1/17 e_{\odot}{e}^{2}  ( {  c_{\odot}}-1) ^{2}  ( c-1  ) ^{2}\cos  ( {  l_{\odot}}+2 g-2 h-2\lambda_{\odot}+4 {  h_{\odot}}  ) -\\
   & 1/17 e_{\odot}{e}^{2}  ( 1+{  c_{\odot}}) ^{2}  ( 1+c  ) ^{2}\cos  ( {  l_{\odot}}+2 g+2 h-2\lambda_{\odot}  ) - \\
   & 1/17 e_{\odot}{e}^{2}  ( {  c_{\odot}}-1  ) ^{2}( 1+c  ) ^{2}\cos  ( -{  l_{\odot}}+2 g+2 h+2 \lambda_{\odot}-4{  h_{\odot}}  ) +\\
& {\frac {7 e_{\odot}{e}^{2}  ( 1+{  c_{\odot}}  ) ^{2}( 1+c  ) ^{2}\cos  ( -{  l_{\odot}}+2 g+2 h-2 \lambda_{\odot}) }{17}}-\\
&{\frac {5   ( {  c_{\odot}}-1  ) ^{2}{e}^{2}( {e_{\odot}}^{2}-2/5  )   ( 1+c  ) ^{2}\cos  ( 2 g+2h+2 \lambda_{\odot}-4 {  h_{\odot}}  ) }{17}}+\\
& {\frac {7 e_{\odot}{e}^{2}( 1+{  c_{\odot}}  ) ^{2}  ( c-1  ) ^{2}\cos  ( 2 g-2 h+2\lambda_{\odot}+{  l_{\odot}}  ) }{17}}-\\
&{\frac {5 {e}^{2}  ( {e_{\odot}}^{2}-2/5)( 1+{  c_{\odot}}  ) ^{2}  ( 1+c  ) ^{2}\cos( 2 g+2 h-2 \lambda_{\odot}  ) }{17}}-\\
&{\frac {5   ( {  c_{\odot}}-1  ) ^{2}  ( c-1  ) ^{2}{e}^{2}  ( {e_{\odot}}^{2}-2/5) \cos  ( 2 g-2 h-2 \lambda_{\odot}+4 {  h_{\odot}}  ) }{17}}-\\
&{\frac {5   ( c-1  ) ^{2}{e}^{2}  ( {e_{\odot}}^{2}-2/5  )( 1+{  c_{\odot}}  ) ^{2}\cos  ( 2 g-2 h+2 \lambda_{\odot}) }{17}}+\\
&{e_{\odot}}^{2}{e}^{2}  ( {  c_{\odot}}-1  ) ^{2}  ( c-1  ) ^{2}\cos  ( -2 {  l_{\odot}}+2 g-2 h-2 \lambda_{\odot}+4 {h_{\odot}}  )+\\
&{e_{\odot}}^{2}{e}^{2}  ( 1+{  c_{\odot}}  ) ^{2}  ( 1+c) ^{2}\cos  ( -2 {  l_{\odot}}+2 g+2 h-2 \lambda_{\odot}  ) \\
&+{e_{\odot}}^{2}{e}^{2}  ( 1+{  c_{\odot}}  ) ^{2}  ( c-1  ) ^{2}\cos  ( 2 {  l_{\odot}}+2 g-2 h+2 \lambda_{\odot}  ) +\\
&{e_{\odot}}^{2}{e}^{2}( {  c_{\odot}}-1  ) ^{2}  ( 1+c  ) ^{2}\cos  ( 2{  l_{\odot}}+2 g+2 h+2 \lambda_{\odot}-4 {  h_{\odot}}  )+\\
&{\frac {108 {e_{\odot}}^{2}  (   ( {c}^{2}+1  ) {{  c_{\odot}}}^{2}+2 {s}^{2}{{  s_{\odot}}}^{2}+{c}^{2}-5/3  )   ( {e}^{2}+2/3  ) \cos  ( 2 {l_{\odot}}  ) }{85}}-\\
&{\frac {18 {e}^{2}{e_{\odot}}^{2}  (   ( {c}^{2}-1  ) {{  c_{\odot}}}^{2}+2 {s}^{2}{{  s_{\odot}}}^{2}+{c}^{2}-1) \cos  ( 2 g+2 {  l_{\odot}}  ) }{17}}+\\
&2 {e}^{2}(   ( {c}^{2}-1  ) {{  c_{\odot}}}^{2}+2 {s}^{2}{{  s_{\odot}}}^{2}-{c}^{2}+1  ) {e_{\odot}}^{2}\cos  ( -2 {  l_{\odot}}-2\lambda+ \\
	& 2 {h_{\odot}}+2 g  ) +\\
&{\frac {14 {e}^{2}  (   ( {c}^{2}-1) {{  c_{\odot}}}^{2}+ 2 {s}^{2}{{  s_{\odot}}}^{2}-{c}^{2}+1  ) e_{\odot}\cos  ( -{  l_{\odot}}-2 \lambda_{\odot}+2 {  h_{\odot}}+2 g  ) }{17}}\\
&+2 {e}^{2}  (   ( {c}^{2}-1  ) {{  c_{\odot}}}^{2}+\\
&{s}^{2}{{s_{\odot}}}^{2}-{c}^{2}+1  ) {e_{\odot}}^{2}\cos  ( 2 {  l_{\odot}}+2 \lambda_{\odot}-2{  h_{\odot}}+2 g  ) -\\
&{\frac {18 {e}^{2}{e_{\odot}}^{2}  (   ( {c}^{2}-1  ) {{  c_{\odot}}}^{2}+2 {s}^{2}{{  s_{\odot}}}^{2}+{c}^{2}-1) \cos  ( 2 g-2 { l_{\odot}}  ) }{17}}-\\
&{\frac {10 {e}^{2} }{17}}  ( {e_{\odot}}^{2}-2/5  )   (   ( {c}^{2}-1  ) {{c_{\odot}}}^{2}+2 {s}^{2}{{  s_{\odot}}}^{2}-{c}^{2}+1  ) \cos  ( 2\lambda_{\odot}- \\
&2 {  h_{\odot}}+2 g  )-\\
&{\frac {12 {e}^{2}  ( {e_{\odot}}^{2}+2/3  )   (   ( {c}^{2}-1  ) {{  c_{\odot}}}^{2}+2 {s}^{2}{{  s_{\odot}}}^{2}+{c}^{2}-1  ) \cos  ( 2 g) }{17}}-\\
&2/17 {e}^{2}  (   ( {c}^{2}-1  ) {{  c_{\odot}}}^{2}+2 {s}^{2}{{  s_{\odot}}}^{2}-{c}^{2}+1  ) e_{\odot}\cos  ( {  l_{\odot}}-2 \lambda_{\odot}+\\
&2{  h_{\odot}}+2 g  ) -\\
& {\frac {12 {e}^{2}e_{\odot}  (   ( {c}^{2}-1) {{  c_{\odot}}}^{2}+2 {s}^{2}{{  s_{\odot}}}^{2}+{c}^{2}-1  )\cos  ( -{  l_{\odot}}+2 g  ) }{17}}+\\
& {\frac {14 {e}^{2}  (( {c}^{2}-1  ) {{  c_{\odot}}}^{2}+2 {s}^{2}{{  s_{\odot}}}^{2}-{c}^{2}+1  ) e_{\odot}\cos  ( {  l_{\odot}}+2 \lambda_{\odot}-2 {  h_{\odot}}+2 g) }{17}}-\\
& {\frac {10 {e}^{2} }{17}}  ( {e_{\odot}}^{2}-2/5  )(   ( {c}^{2}-1  ) {{  c_{\odot}}}^{2}+2 {s}^{2}{{  s_{\odot}}}^{2}-{c}^{2}+1  ) \cos  ( -2 \lambda_{\odot}+\\
&2 {  h_{\odot}}+2 g) - {\frac {12 {e}^{2}e_{\odot}  (   ( {c}^{2}-1  ){{  c_{\odot}}}^{2}+2 {s}^{2}{{  s_{\odot}}}^{2}+{c}^{2}-1  ) \cos  ({  l_{\odot}}+2 g  ) }{17}}-\\
& 2/17 {e}^{2}  (   ( {c}^{2}-1) {{  c_{\odot}}}^{2}+2 {s}^{2}{{  s_{\odot}}}^{2}-{c}^{2}+1  ) e_{\odot}\cos  ( -{  l_{\odot}}+2 \lambda_{\odot}-\\
&2 {  h_{\odot}}+2 g  ) +\\
& {\frac {28e_{\odot}s{  s_{\odot}} {e}^{2}  ( {  c_{\odot}}-1  )   ( c-1  )\cos  ( 2 g-2 \lambda_{\odot}+3 {  h_{\odot}}-{  l_{\odot}}-h  ) }{17}}+\\
& {\frac {216 {  c_{\odot}} {  s_{\odot}} {e_{\odot}}^{2}sc  ( {e}^{2}+2/3  )\cos  ( h-{  h_{\odot}}+2 {  l_{\odot}}  ) }{85}}+ \\
& {\frac {28 e_{\odot}s{s_{\odot}} {e}^{2}  ( 1+{  c_{\odot}}  )   ( c-1  ) \cos( {  l_{\odot}}+2 g+2 \lambda_{\odot}-{  h_{\odot}}-h  ) }{17}}+\\
& {\frac {28e_{\odot}s{  s_{\odot}} {e}^{2}  ( 1+{  c_{\odot}}  )   ( 1+c  )\cos  ( 2 g-2 \lambda_{\odot}+{  h_{\odot}}-{  l_{\odot}}+h  ) }{17}}-\\
& {\frac{36 {e_{\odot}}^{2}s{  s_{\odot}} {e}^{2}{  c_{\odot}}   ( c-1  ) \cos( 2 g-h+{  h_{\odot}}-2 {  l_{\odot}}  ) }{17}}-\\
& {\frac {36 {e_{\odot}}^{2}s{  s_{\odot}} {e}^{2}{  c_{\odot}}   ( 1+c  ) \cos  ( 2 g-{h_{\odot}}+h-2 {  l_{\odot}}  ) }{17}}-\\
& {\frac {24 e_{\odot}{  c_{\odot}} s{  s_{\odot}} {e}^{2}  ( c-1  ) \cos  ( 2 g+{  h_{\odot}}-h+{  l_{\odot}}  )}{17}}+\\
& {\frac {144 {  c_{\odot}} {  s_{\odot}} e_{\odot}sc  ( {e}^{2}+2/3) \cos  ( -{  h_{\odot}}+h+{  l_{\odot}}  ) }{85}}+\\
&{\frac {28 e_{\odot}s{  s_{\odot}} {e}^{2}  ( {  c_{\odot}}-1  )   ( 1+c  ) \cos( 2 g+2 \lambda_{\odot}-3 {  h_{\odot}}+{  l_{\odot}}+h  ) }{17}}-\\
&{\frac {20 {e}^{2}  ( {e_{\odot}}^{2}-2/5  )   ( 1+{  c_{\odot}}  ) {s_{\odot}} s  ( 1+c  ) \cos  ( 2 g-2 \lambda_{\odot}+{  h_{\odot}}+h) }{17}}-\\
& {\frac {24 e_{\odot}{  c_{\odot}} s{  s_{\odot}} {e}^{2}  ( 1+c) \cos  ( 2 g-{  h_{\odot}}+h-{  l_{\odot}}  ) }{17}}-\\
& {\frac {24 e_{\odot}{  c_{\odot}} s{  s_{\odot}} {e}^{2}  ( 1+c  ) \cos  ( {l_{\odot}}+2 g-{  h_{\odot}}+h  ) }{17}}-\\
& {\frac {4 e_{\odot}s{  s_{\odot}} {e}^{2}( 1+{  c_{\odot}}  )   ( 1+c  ) \cos  ( 2 g-2\lambda_{\odot}+{  h_{\odot}}+{  l_{\odot}}+h  ) }{17}}+\\
&{\frac {216 {  c_{\odot}} {s_{\odot}} {e_{\odot}}^{2}sc  ( {e}^{2}+2/3  ) \cos  ( -{  h_{\odot}}+h-2 {  l_{\odot}}  ) }{85}}-\\
& {\frac {24 {e}^{2}  ( {e_{\odot}}^{2}+2/3) {  c_{\odot}} {  s_{\odot}} s  ( 1+c  ) \cos  ( 2 g-{h_{\odot}}+h  ) }{17}}-\\
& {\frac {24 e_{\odot}{  c_{\odot}} s{  s_{\odot}} {e}^{2}( c-1  ) \cos  ( 2 g+{  h_{\odot}}-h-{  l_{\odot}}  ) }{17}}-\\
& {\frac {4 e_{\odot}s{  s_{\odot}} {e}^{2}  ( {  c_{\odot}}-1  )   ( c-1) \cos  ( 2 g-2 \lambda_{\odot}+3 {  h_{\odot}}+{  l_{\odot}}-h  ) }{17}}+\\
& {\frac {144 {  c_{\odot}} {  s_{\odot}} e_{\odot}sc  ( {e}^{2}+2/3  )\cos  ( -{  l_{\odot}}-{  h_{\odot}}+h  ) }{85}}-\\
& {\frac {36 {e_{\odot}}^{2}s{s_{\odot}} {e}^{2}{  c_{\odot}}   ( c-1  ) \cos  ( 2 g+{  h_{\odot}}-h+2 {  l_{\odot}}  ) }{17}}-\\
& {\frac {36 {e_{\odot}}^{2}s{  s_{\odot}} {e}^{2}{c_{\odot}}   ( 1+c  ) \cos  ( 2 g-{  h_{\odot}}+h+2 {  l_{\odot}}) }{17}}-\\
& {\frac {4 e_{\odot}s{  s_{\odot}} {e}^{2}  ( {  c_{\odot}}-1)   ( 1+c  ) \cos  ( 2 g+2 \lambda_{\odot}-3 {  h_{\odot}}-{l_{\odot}}+h  ) }{17}}-\\
& {\frac {4 e_{\odot}s{  s_{\odot}} {e}^{2}  ( 1+{c_{\odot}}  )   ( c-1  ) \cos  ( -{  l_{\odot}}+2 g+2 \lambda_{\odot}-{  h_{\odot}}-h  ) }{17}}\\
&+4 {e_{\odot}}^{2}s{  s_{\odot}} {e}^{2}  ( {  c_{\odot}}-1  )   ( c-1  ) \cos  ( 2 g-2 \lambda_{\odot}+3 {  h_{\odot}}-2 {  l_{\odot}}-h  )\\
&+4 {e_{\odot}}^{2}s{  s_{\odot}} {e}^{2}  ( 1+{  c_{\odot}})   ( c-1  ) \cos  ( 2 {  l_{\odot}}+2 g+2 \lambda_{\odot}-{	h_{\odot}}-h  ) \\
&+4 {e_{\odot}}^{2}s{  s_{\odot}} {e}^{2}  ( {  c_{\odot}}-1)   ( 1+c  ) \cos  ( 2 {  l_{\odot}}+2 g+2 \lambda_{\odot}-3{  h_{\odot}}+h \\
&+4 {e_{\odot}}^{2}s{  s_{\odot}} {e}^{2}  ( 1+{  c_{\odot}})   ( 1+c  ) \cos  ( -2 {  l_{\odot}}+2 g-2 \lambda_{\odot}+{  h_{\odot}}+h  ) \\
&-{\frac {6 e_{\odot}{e}^{2}  ( {  c_{\odot}}-1  )( 1+{  c_{\odot}}  )   ( 1+c  ) ^{2}\cos  ( 2 g+2h-2 {  h_{\odot}}-{  l_{\odot}}  ) }{17}}\\
&-{\frac {9 {e_{\odot}}^{2}{e}^{2}( {  c_{\odot}}-1  )   ( 1+{  c_{\odot}}  )   ( c-1) ^{2}\cos  ( 2 g-2 h+2 {  h_{\odot}}-2 {  l_{\odot}}  ) }{17}}\\
&-6/5   ( {  c_{\odot}}-1  ) ^{2}  ( c-1  ) {e_{\odot}}^{2}( 1+c  )   ( {e}^{2}+2/3  ) \cos  ( 2 {  l_{\odot}}+2 \lambda_{\odot}-\\
&4 {  h_{\odot}}+2 h  )-\\
&{\frac {6 e_{\odot}{e}^{2}  ( {c_{\odot}}-1  )   ( 1+{  c_{\odot}}  )   ( 1+c  ) ^{2}\cos  ( 2 g+2 h-2 {  h_{\odot}}+{  l_{\odot}}  ) }{17}}-\\
&{\frac {42( {  c_{\odot}}-1  ) ^{2}  ( c-1  ) e_{\odot}  ( 1+c)   ( {e}^{2}+2/3  ) \cos  ( {  l_{\odot}}+2 \lambda_{\odot}-4{  h_{\odot}}+2 h  ) }{85}}-\\
&{\frac {9 {e_{\odot}}^{2}{e}^{2}  ( {c_{\odot}}-1  )   ( 1+{  c_{\odot}}  )   ( 1+c  ) ^{2}\cos( 2 g+2 h-2 {  h_{\odot}}-2 {  l_{\odot}}  ) }{17}}-\\
&{\frac {6 e_{\odot}{e}^{2}  ( {  c_{\odot}}-1  )   ( 1+{  c_{\odot}}  )   ( c-1  ) ^{2}\cos  ( -{  l_{\odot}}+2 g-2 h+2 {  h_{\odot}}  ) }{17}}+\\
&{\frac {6 }{17}}  ( {  c_{\odot}}-1  ) ^{2}  ( c-1  )( {e_{\odot}}^{2}-2/5  )   ( 1+c  )   ( {e}^{2}+2/3) \cos  ( 2 \lambda_{\odot}-\\
&4 {  h_{\odot}}+2 h  ) +\\
&{\frac {6   ( {  c_{\odot}}-1  ) ^{2}  ( c-1  ) e_{\odot}  (1+c  )   ( {e}^{2}+2/3  ) \cos  ( -{  l_{\odot}}+2\lambda_{\odot}-4 {  h_{\odot}}+2 h  ) }{85}}-\\
&{\frac {6 e_{\odot}{e}^{2}  ( {c_{\odot}}-1  )   ( 1+{  c_{\odot}}  )   ( c-1  ) ^{2}\cos  ( {  l_{\odot}}+2 g-2 h+2 {  h_{\odot}}  ) }{17}}-\\
&{\frac {9 {e_{\odot}}^{2}{e}^{2}  ( {  c_{\odot}}-1  )   ( 1+{  c_{\odot}}  )( 1+c  ) ^{2}\cos  ( 2 g+2 h-2 {  h_{\odot}}+2 {  l_{\odot}}) }{17}}-\\
&{\frac {9 {e_{\odot}}^{2}{e}^{2}  ( {  c_{\odot}}-1  )( 1+{  c_{\odot}}  )   ( c-1  ) ^{2}\cos  ( 2 g-2h+2 {  h_{\odot}}+2 {  l_{\odot}}  ) }{17}}-\\
&{\frac {168 c  ( 1+{c_{\odot}}  ) s  ( {e}^{2}+2/3  ) {  s_{\odot}} e_{\odot}\cos  ( -{l_{\odot}}-2 \lambda_{\odot}+{  h_{\odot}}+h  ) }{85}}-\\
&{\frac {24 c  ( {c_{\odot}}-1  ) s  ( {e}^{2}+2/3  ) {  s_{\odot}} {e_{\odot}}^{2}\cos( 2 {  l_{\odot}}+2 \lambda_{\odot}-3 {  h_{\odot}}+h  ) }{5}}+\\
&{\frac {24c  ( {  c_{\odot}}-1  ) s  ( {e}^{2}+2/3  ) {  s_{\odot}} e_{\odot}\cos  ( 2 \lambda_{\odot}-3 {  h_{\odot}}-{  l_{\odot}}+h  ) }{85}}+\\
&{\frac {24 c  ( 1+{  c_{\odot}}  ) s  ( {e_{\odot}}^{2}-2/5  )   ( {e}^{2}+2/3  ) {  s_{\odot}} \cos  ( -2 \lambda_{\odot}+{  h_{\odot}}+h) }{17}}+\\
&{\frac {144 c{  c_{\odot}} s  ( {e}^{2}+2/3  ) {s_{\odot}}   ( {e_{\odot}}^{2}+2/3  ) \cos  ( -{  h_{\odot}}+h  )}{85}}+\\
&{\frac {24 c  ( 1+{  c_{\odot}}  ) s  ( {e}^{2}+2/3) {  s_{\odot}} e_{\odot}\cos  ( {  l_{\odot}}-2 \lambda_{\odot}+{  h_{\odot}}+h) }{85}}-\\
&{\frac {24 {  c_{\odot}}   ( c-1  ) s{e}^{2}{s_{\odot}}   ( {e_{\odot}}^{2}+2/3  ) \cos  ( 2 g+{  h_{\odot}}-h) }{17}}-\\
&{\frac {168 c  ( {  c_{\odot}}-1  ) s  ( {e}^{2}+2/3  ) {  s_{\odot}} e_{\odot}\cos  ( 2 \lambda_{\odot}-3 {  h_{\odot}}+{  l_{\odot}} +h  ) }{85}}+\\
&{\frac {24 c  ( {  c_{\odot}}-1  ) s  ( {e_{\odot}}^{2}-2/5  )   ( {e}^{2}+2/3  ) {  s_{\odot}} \cos  ( 2\lambda_{\odot}-3 {  h_{\odot}}+h  ) }{17}}-\\
&{\frac {24 c  ( 1+{  c_{\odot}}) s  ( {e}^{2}+2/3  ) {  s_{\odot}} {e_{\odot}}^{2}\cos  ( -2{  l_{\odot}}-2 \lambda_{\odot}+{  h_{\odot}}+h  ) }{5}}+\\
&{\frac {6 }{17}}  ( 1+{c_{\odot}}  ) ^{2}  ( c-1  )   ( 1+c  )   ( {e_{\odot}}^{2}-2/5  )   ( {e}^{2}+2/3  ) \cos  ( 2 h-\\
& 2\lambda_{\odot}  ) +{\frac {6 }{85}}   ( 1+{  c_{\odot}}  ) ^{2}( c-1  )   ( 1+c  )   ( {e}^{2}+2/3  ) e_{\odot}\cos  ( {  l_{\odot}}+\\
& 2 h- 2 \lambda_{\odot}  )-6/5   ( 1+{c_{\odot}}  ) ^{2}  ( c-1  )   ( 1+c  )   ( {e}^{2}+2/3  ) {e_{\odot}}^{2}\times \\
&\cos  ( -2 {  l_{\odot}}+2 h-2 \lambda_{\odot}) -\\
&{\frac {42   ( 1+{  c_{\odot}}  ) ^{2}  ( c-1		)   ( 1+c  )   ( {e}^{2}+2/3  ) e_{\odot}\cos  (-{  l_{\odot}}+2 h-2 \lambda_{\odot}  ) }{85}}+\\
&{\frac{72}{85}}  (( {c}^{2}+1  ) {{  c_{\odot}}}^{2}+2 {s}^{2}{{  s_{\odot}}}^{2}+{c}^{2}-5/3  )   ( e_{\odot}\cos  ( {  l_{\odot}}  ) +1/2 {e_{\odot}}^{2}+\\
& 1/3  )   ( {e}^{2}+2/3  ) -{\frac{20}{17}}  ( {  c_{\odot}}-1  )   ( 1+c  ) s  ( {e_{\odot}}^{2}-2/5  ) {e}^{2}{s_{\odot}} \times \\
& \cos  ( 2 g+2 \lambda_{\odot}-3 {  h_{\odot}}+h  ) +\\
&{\frac{54}{85}}  ( {  c_{\odot}}-1  )   ( 1+{  c_{\odot}}  )   ( c-1)   ( 1+c  )   ( {e}^{2}+2/3  ) {e_{\odot}}^{2} \times \\
& \cos( -2 {  h_{\odot}}+2 h-2 {  l_{\odot}}  ) -\\
&{\frac{6}{17}}  ( {c_{\odot}}-1  )   ( 1+{  c_{\odot}}  )   ( c-1  ) ^{2}{e}^{2}  ( {e_{\odot}}^{2}+2/3  ) \cos  ( 2 g-2 h+\\
& 2 {  h_{\odot}}) -{\frac{6}{17}}  ( {  c_{\odot}}-1  )   ( 1+{  c_{\odot}})   ( 1+c  ) ^{2}{e}^{2}  ( {e_{\odot}}^{2}+2/3  )\cos  ( 2 g+\\
& 2 h-2 {  h_{\odot}}  ) -{\frac{12}{5}}  (( {c}^{2}+1  ) {{  c_{\odot}}}^{2}+2 {s}^{2}{{  s_{\odot}}}^{2}-{c}^{2}-1  )   ( {e}^{2}+2/3  ) \times \\
& {e_{\odot}}^{2}\cos  ( -2 {l_{\odot}}-2 \lambda_{\odot}+2 {  h_{\odot}}  ) -\\
&{\frac{84}{85}}  (   ( {c}^{2}+1  ) {{  c_{\odot}}}^{2}+2 {s}^{2}{{  s_{\odot}}}^{2}-{c}^{2}-1)   ( {e}^{2}+2/3  ) \times \\
& e_{\odot}\cos  ( -{  l_{\odot}}-2 \lambda_{\odot}+2 {  h_{\odot}}  ) +\\
&{\frac{12}{85}}  (   ( {c}^{2}+1) {{  c_{\odot}}}^{2}+2 {s}^{2}{{  s_{\odot}}}^{2}-{c}^{2}-1  )( {e}^{2}+2/3  ) e_{\odot}\cos  ( {  l_{\odot}}-\\
&2 \lambda_{\odot}+2 {h_{\odot}}  ) +\\
&{\frac{12}{17}}  (   ( {c}^{2}+1  ) {{  c_{\odot}}}^{2}+2 {s}^{2}{{  s_{\odot}}}^{2}-{c}^{2}-1  )   ( {e_{\odot}}^{2}-2/5)   ( {e}^{2}+\\
&2/3  ) \cos  ( -2 \lambda_{\odot}+2 {h_{\odot}}  ) +\\
&{\frac{54}{85}}  ( {  c_{\odot}}-1  )   ( 1+{c_{\odot}}  )   ( c-1  )   ( 1+c  )   ( {e}^{2}+2/3  ) {e_{\odot}}^{2}\cos  ( -2 {  h_{\odot}}+\\
&2 h+2 {  l_{\odot}}  ) +\\
&{\frac{36}{85}}  ( {  c_{\odot}}-1  )   ( 1+{  c_{\odot}}  )( c-1  )   ( 1+c  )   ( {e}^{2}+2/3  ) e_{\odot}\cos  ( -2 {  h_{\odot}}+\\
&2 h-{  l_{\odot}}  ) +\\
&{\frac{36}{85}}( {  c_{\odot}}-1  )   ( 1+{  c_{\odot}}  )   ( c-1)   ( 1+c  )   ( {e}^{2}+2/3  ) e_{\odot}\cos  (-2 {  h_{\odot}}+\\
&2 h+{  l_{\odot}}  ) -\\
&{\frac{20}{17}}  ( 1+{  c_{\odot}})   ( c-1  ) s  ( {e_{\odot}}^{2}-2/5  ) {e}^{2}{s_{\odot}} \cos  ( 2 g+2 \lambda_{\odot}-{  h_{\odot}}-h  ) -\\
&{\frac{20}{17}}( {  c_{\odot}}-1  )   ( c-1  ) s  ( {e_{\odot}}^{2}-2/5) {e}^{2}{  s_{\odot}} \cos  ( 2 g-2 \lambda_{\odot}+3 {  h_{\odot}}-h) +\\
&{\frac{36}{85}}  ( {  c_{\odot}}-1  )   ( 1+{  c_{\odot}})   ( c-1  )   ( 1+c  )   ( {e}^{2}+2/3)   ( {e_{\odot}}^{2}+\\
&2/3  ) \cos  ( -2 {  h_{\odot}}+2 h)   ),
\end{align*}
where $\lambda_{\odot}=l_{\odot}+g_{\odot}+h_{\odot}=$ is the mean longitude of the third body (Sun), $c_{\odot}=\cos(i_{\odot})$ and $s_{\odot}=\sin(i_{\odot})$.

\section{Acknowledgments}

The authors wish to express their appreciation for the support provided by grants 307724/2017-4, 303102/2019-5 from the National Council for Scientific and Technological Development (CNPq); grants 2016/24561-0 from S\~{a}o Paulo Research Foundation (FAPESP) and the financial support from the Coordination for the Improvement of Higher Education Personnel (CAPES).



\end{document}